\documentclass[sigconf,authorversion]{acmart}
\renewcommand\footnotetextcopyrightpermission[1]{} %

\acmConference{Accepted to SIGMOD'26}
\acmBooktitle{Bengaluru, India}
\usepackage[colorinlistoftodos,prependcaption,textsize=normalsize]{todonotes}

\usepackage{multirow}
\usepackage{makecell}

\newcommand{\sys}{P-MOSS}
\usepackage{enumitem}
\usepackage{amsmath}

\usepackage{booktabs}  
\usepackage{subcaption}
\newcommand{\labeltext}[2]{%
  \@bsphack
  \csname phantomsection\endcsname 
  \def\@currentlabel{#1}{\label{#2}}%
  \@esphack
}
\usepackage{cleveref}
\usepackage{caption} 

\usepackage{tikz}
\usepackage{xcolor}
\usepackage{enumitem}
\newcounter{goal}

\newcounter{block}

\definecolor{firebrick}{HTML}{B22222} 
\definecolor{clightgray}{rgb}{0.83, 0.83, 0.83}

\newcommand{\styledtextcircledo}[3]{
  \tikz[baseline={(char.base)},outer sep=0pt]{
    \node[shape=circle, draw=#1, 
    line width=0.1mm, 
    fill=#2, 
    text=white,                
    minimum size=4pt,
    inner sep=0.2pt] (char) {{\small{#3}}};
  }
}

\newcommand{\ciro}[1]{
  \styledtextcircledo{black}{firebrick}{\textbf{#1}}
}
\newcommand{\ciroo}[1]{
  \styledtextcircledo{black}{blue}{\textbf{#1}}
}
\definecolor{customgreen}{HTML}{339933}
\newcommand{\cirooo}[1]{
  \styledtextcircledo{black}{customgreen}{\textbf{#1}}
}
\definecolor{lightpink}{RGB}{255, 182, 193}
\newcommand{\lpbullet}{%
    \tikz{\fill[lightpink] (0,0) circle (0.1cm); \draw (0,0) circle (0.1cm);}
}
\definecolor{lightsalmon}{RGB}{255, 160, 122}
\newcommand{\lss}{%
    \tikz{\draw[black, fill=lightsalmon] (0,0) rectangle (0.2,0.2);}%
}

\usepackage{bm}
\newcommand{\Prob}{\mathbb{P}}
\begin{document}

\title{\sys{}: Scheduling Main-Memory Indexes Over NUMA Servers Using Next Token Prediction}
\markboth{Accepted in SIGMOD ’26}{Accepted in SIGMOD ’26}
\author{Yeasir Rayhan}
\orcid{0000-0003-0326-2965}
\affiliation{%
  \institution{Purdue University}
  \country{West Lafayette, IN, USA}}
\email{yrayhan@purdue.edu}

\author{Walid G. Aref}
\orcid{0000-0001-8169-7775}
\affiliation{%
  \institution{Purdue University}
  \country{West Lafayette, IN, USA}}
\email{aref@purdue.edu}

 \renewcommand{\shortauthors}{Yeasir Rayhan and Walid G. Aref}

\begin{abstract}
Ever since the Dennard scaling broke down in the early 2000s and the frequency of the CPUs stalled, vendors have started to increase the core count in each CPU chip at the expense of introducing heterogeneity, thus ushering the era of NUMA and Chiplet processors. Since then, the heterogeneity in the design space of hardware has only increased to the point that DBMS performance may vary significantly up to an order of magnitude in modern servers. An important factor that affects performance includes the location of the logical cores where the DBMS queries execute, and the location where the data resides. 
This paper introduces \sys{}, a learned spatial scheduling framework that schedules query execution to specific logical cores, and co-locates data on the corresponding NUMA node. For cross-hardware and workload adaptability, \sys{} leverages core principles from Large Language Models, such as Next Token prediction, Generative Pre-training, and Fine-tuning. 
In the spirit of hardware-software synergy, \sys{} guides its scheduling decision solely based on the low-level hardware statistics collected from the hardware Performance Monitoring Unit with the aid of a Decision Transformer. Experimental evaluation is performed in the context of the B$^+$-Tree index. Performance results demonstrate that \sys{} offers an improvement of up to $6\times$ over traditional schedules in terms of query throughput. 
\end{abstract}



\keywords{Spatial Scheduling; NUMA \& Chiplet Processors; Hardware Performance Monitoring Unit (PMU); Offline RL; Next Token Prediction}


\maketitle

\markboth{Accepted in SIGMOD ’26}{Accepted in SIGMOD ’26}

\section{Introduction}

\begin{figure}[tbp]
    \captionsetup{belowskip=-15pt}
    \captionsetup{aboveskip=-1pt}
    \centering
    \includegraphics[width=\columnwidth]{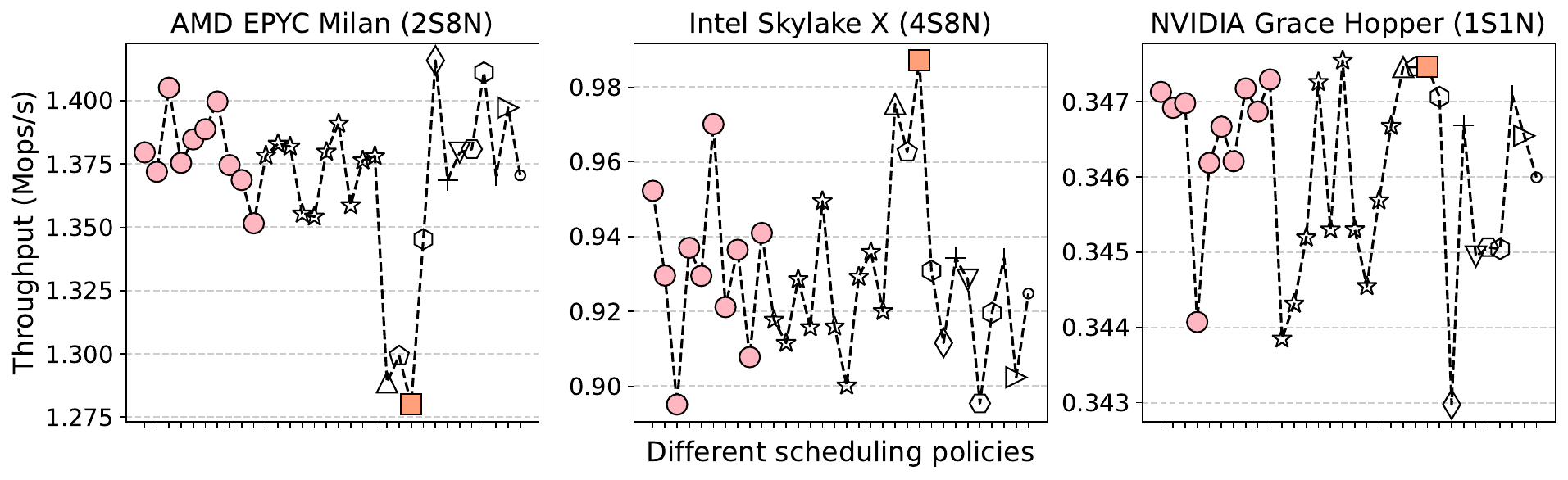}
    \caption{The performance of a B$^+$-Tree with 30M initial records under different spatial scheduling policies.
    }
    \label{fig:motivating_img}
\end{figure}

Due to the hardware heterogeneity of modern multi-core NUMA and Chiplet servers, the data partitioning strategy and the core scheduling policy of a main-memory index can significantly dictate the index's performance. The data partitioning strategy decides the location, i.e., the NUMA node of an index partition. The core scheduling policy decides the hardware core, i.e., the core ID responsible for executing an incoming query targeting an index partition. We refer to this combination of data placement strategy and core scheduling policy as spatial scheduling~\cite{GicevaARH14}. The performance gap of a B$^+$-Tree under different scheduling policies can differ by up to 5.91$\times$ for different NUMA machines (See~\S\ref{sec:eval}). Figure~\ref{fig:motivating_img} 
illustrates
the performance of a main-memory B$^+$-Tree~\cite{Comer79} in 
three different multi-core servers under the 50\% read-write YCSB~\cite{CooperSTRS10} workload. Each data point refers to a distinct spatial scheduling policy of the B$^+$-Tree. Data points with the same marker (cf., \lpbullet, \lss~for the AMD server) denote the same data partitioning strategy. It is evident that the same B$^+$-Tree can exhibit different performance under different data partitioning strategies. Even two policies with the same data placement but a different core scheduling policy can yield very different performance (cf. \lpbullet markers~for the Intel server). 
Moreover, no one policy performs optimally across all the servers (cf. \lss ~marker across the AMD and  Intel 
servers).

In this paper, we focus on the spatial scheduling problem of a main-memory index, and introduce \textbf{\sys{}}, a learned \underline{P}erformance \underline{MO}nitoring Unit (PMU)-driven \underline{S}patial Query \underline{S}cheduling framework. It utilizes the \textbf{Next Token Prediction} paradigm (NTP, for short)  via a Decision Transformer to improve the query execution performance of main-memory indexes in NUMA and Chiplet servers. To handle larger indexes, \sys{} \textit{logically} partitions each index into multiple index \textit{slices} based on the index key. Each slice corresponds to a specific key range. These index slices serve as the unit of spatial scheduling. \sys{} addresses the following question: \textbf{Given a main-memory index that is logically partitioned into multiple index slices on a NUMA or Chiplet server, how to find the best possible mapping for each index slice to a CPU core and a NUMA node, i.e., memory controller, such that the index performance, i.e., its throughput, is maximized.} To the best of our knowledge, \sys{} is the first to address the problem of spatial query scheduling for main-memory indexes in modern multi-core NUMA and Chiplet servers. Prior works on query scheduling, e.g.,~\cite{PorobicPBTA12,PorobicLTA14,LeisBK014,WagnerK021,PsaroudakisSMSA15,PsaroudakisSMSA16,MaoSVMA19,SabekUK22}, draw upon the temporal aspect of query scheduling, \textit{partially} (~\cite{PorobicPBTA12,PorobicLTA14,PsaroudakisSMSA15,PsaroudakisSMSA16} only consider data placement in the context of relational tables) or \textit{completely} ignoring its spatial aspect, and are orthogonal to the \sys's approach introduced in this paper. 

\begin{figure}[htbp]
    \captionsetup{belowskip=-10pt}
    \captionsetup{aboveskip=-2pt}
    \centering
    \includegraphics[width=\columnwidth]{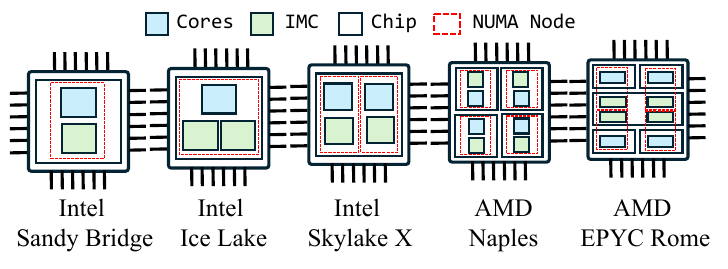}
    \caption{Different socket layouts of NUMA, Chiplet servers.}
    \label{fig:numa_eval}
\end{figure}

Designing efficient schedules for main-memory indexes is particularly challenging due to the increasing complexity of modern hardware and query workloads. The current hardware landscape is highly diverse, with multiple CPU vendors, e.g., Intel, AMD, NVIDIA, and Amazon, as well as a wide range of NUMA and Chiplet architectures (cf. Figure~\ref{fig:numa_eval}). 
Cloud providers,
e.g.,
Amazon 
EC2~\cite{amazon-ec2}, provide as many as 750 instance choices equipped with a diverse range of processors, memory, storage and networks. These servers 
have
distinct characteristics 
in 
topology and 
performance. 
For example, 
for a 4-Socket Intel Skylake X~\cite{intel_skx} and a 2-Socket AMD Milan~\cite{amd_epyc}, the inter-core latencies within the socket can vary by up to 1.1$\times$ and 4$\times$, respectively. This can increase by up to 3$\times$ and 10$\times$ across sockets.\footnote{All the numbers are generated on our testbed servers (See \S~\ref{sec:eval}).} Moreover, the number of CPU cores per socket has increased 
over the years. Modern Chiplet processors,
e.g.,
Intel Sierra Forest-SP~\cite{intel_sierra_forest} and AMD EPYC 9755~\cite{amd_epyc_9755} offer up to 144 and 128 cores, respectively. Even for servers with a high core count that do not employ the Chiplet architecture, e.g., a 72-core NVIDIA GH200 Grace Hopper~\cite{nvidia_gh}, the inter-core latency between distant cores can vary up to $1.5\times$. This makes
efficient spatial scheduling 
critical for high-performance indexes. 
Moreover,
indexes need to support a wide range of workload 
patterns, e.g., read-heavy, write-heavy, scan-intensive operations, with distinct data distributions. \sys{} adapts not only to the diverse CPU vendors and the ever evolving NUMA, Chiplet architectures, but also efficiently handles 
a wide range of 
query workloads. 

Modern Large Language Models (LLM)~\cite{BrownMRSKDNSSAA20,Radford2019LanguageMA,Radford2018ImprovingLU,abs-2302-13971,abs-2303-08774,abs-2401-02954} are highly effective at adapting to diverse tasks and contexts. They are highly complex, built upon key components, e.g., the Transformer architecture~\cite{VaswaniSPUJGKP17}, Generative Pre-training~\cite{Radford2019LanguageMA}, Supervised Fine-tuning (SFT)~\cite{Radford2019LanguageMA}, and Reinforcement Learning from Human Feedback (RLHF)~\cite{lambertrlhf}. Despite this complexity, a simple yet elegant idea underpins these building blocks, and remains at the core of modern LLMs, i.e., 
Next Token Prediction (NTP). 
Given a sequence of tokens, i.e., words, sub-words, or letters, NTP refers to the task of estimating the probability of the next token. \sys{} formulates the problem of spatial query scheduling as an
NTP 
task. Given the scheduling decisions for all previous $\langle i-1\rangle$ index slices, \sys{} predicts the next core to place the $i$-th index slice. 
To effectively solve the NTP task for index scheduling, \sys{} adopts Reinforcement Learning. 

\sys{} adopts foundational concepts from LLM to solve the index scheduling task. The GPT architecture serves as the backbone of \sys{}'s model architecture. Inspired by the two-phase training process of LLMs, \sys{} also trains the GPT architecture in  a pre-training phase followed by a post-training phase. 
Pre-training aims to install the knowledge of the optimization landscape for the scheduling task in \sys{}, while  post-training aligns \sys{} to the unique characteristics of the target hardware and workload. During pre-training, \sys{} trains the GPT on a large and diverse dataset encompassing scheduling policies across various hardware 
vendors, NUMA, Chiplet architectures, and query workloads. 
By training on diverse datasets, \sys{} learns a model that can identify generalizable patterns of efficient scheduling across different hardware 
and workload patterns. In the post-training stage, \sys{} adapts the trained GPT from the pre-training phase to the target hardware and the target workloads. It  incrementally trains the GPT on data collected from the target hardware and the target workload.

During both training stages, \sys{} adopts {\em Offline Reinforcement Learning} to learn the scheduling policy of a main-memory index. By adopting the offline RL scheme over the traditional RL, \sys{} decouples the learning process from the DBMS kernel. This ensures that the DBMS does not under-perform during the ML agent's learning cycle. 
Once the scheduling policy of each index slice is learned, \sys{} enforces the learned policy by localizing  query execution to the specific cores and distributing the index slices over the corresponding NUMA nodes. A noteworthy aspect of \sys{} is that its learning cycle is solely guided by the hardware performance statistics sampled 
by the {\em Performance Monitoring Unit (PMU)}~\cite{pmuIntel,pmuAMD,pmuARM} of the processor, without any software-level bookkeeping. These statistics are the hardware traces left behind by the executed queries at different parts of the hardware, e.g., cores, caches, memory controllers. The statistics collected from the hardware PMU registers equip \sys{} with the necessary abstraction so that it can be trained on diverse hardware and workload configurations during the pre-training phase. This 
also allows \sys{} to adapt to the target environment during post-training.

The main contribution of this paper is an architectural blueprint for a framework inspired by the foundational concepts from LLMs. \sys{} requires no bookkeeping, and is adaptable across diverse hardware configurations and query workload patterns. More specifically, the contributions of this paper are as follows:

\begin{itemize}[leftmargin=*]
    \item We introduce {\em \sys{}, a learned 
    query scheduling} framework over a main-memory index
    that prioritizes the spatial aspect of query scheduling, i.e., query execution and data placement, 
    at the logical core and NUMA node, i.e., memory controller levels. 
    
    \item We develop DBMS kernel optimizations that are 
    solely guided by the {\em low-level hardware statistics}  collected by the Performance Monitoring Units without any software-level bookkeeping. 
    
    \item We formulate
    the problem of query scheduling as a Next Token Prediction task 
    and adopt {\em Offline RL} to solve it, 
    enabling 
    learning without requiring any online interaction with the DBMS.     
    
    \item We test \sys{} on a main-memory B$^+$-Tree over a 
    range of 
    CPUs (Intel, AMD, ARM, IBM), NUMA and Chiplet hardware (1-4 NUMA Sockets, 2-8 NUMA Nodes), query workload, and have 
    up to
    $6\times$ throughput improvement over traditional scheduling.
\end{itemize}

\section{Related Work}
\noindent\textbf{Temporal Query Scheduling} decides the sequence in which queries are executed on a single core, once the queries are assigned to it. In contrast, spatial scheduling operates at a higher level by deciding both the core assignment and the corresponding data placement. Extensive body of work focuses on temporal query scheduling, e.g., heuristic-driven~\cite{PatelDZPZSMS18,LeisBK014,WagnerK021} and learned~\cite{SabekUK22,MaoSVMA19} strategies. \sys{} employs a First Come First Serve strategy for temporal scheduling on each core, though this can be replaced by any of the aforementioned methods. 

\noindent\textbf{NUMA-awareness in Main Memory DBMSs} 
includes stand-alone NUMA-aware query operators, synchronization, and scheduling strategies. Stream join~\cite{TeubnerM11}, sort-merge join~\cite{AlbutiuKN12,LiPMRL13}, hash join~\cite{BalkesenATO13,LangLANK13}, radix join~\cite{SchuhCD16} are  examples of  existing work on standalone NUMA-aware query operators. 
Other works in NUMA delve into the synchronization primitives, e.g., designing NUMA-friendly locks~\cite{ChabbiFM15,LoziD0LM16,KashyapMK17} and concurrent data structures~\cite{CalciuSBA17}. The NUMA-related studies 
closest 
to \sys{} are NUMA-aware query scheduling~\cite{PorobicPBTA12,PorobicLTA14,LeisBK014,WagnerK021,PsaroudakisSMSA15,PsaroudakisSMSA16}. Morsel-driven query scheduling 
of  HyPer~\cite{LeisBK014} and Umbra~\cite{WagnerK021} 
stores relations as morsels (data fragments), and uniformly distributes them over NUMA sockets. Each worker thread operates on morsels local to it, and writes the results to the local NUMA socket. SAP HANA's  NUMA-aware query scheduling~\cite{PsaroudakisSMSA15,PsaroudakisSMSA16} adaptively partitions tables, and locates them in appropriate NUMA sockets to balance the load across NUMA sockets. In contrast to \sys{}, these works focus on relational tables, and do not consider query scheduling in main-memory indexes. 


\noindent\textbf{Clustering and Declustering} tree index nodes, e.g., B$^+$-Tree~\cite{PramanikK90,SeegerL91} and R-Tree~\cite{KoudasFK96,DiwanRSS96,KamelF92} improve query performance. These  works focus on disk-based indexes and older machine architectures. Buzzard~\cite{MaasKHL13} applies a heuristic-based adaptive data partitioning scheme to distribute a prefix-tree-based index across the NUMA system to optimize remote memory accesses. For robust performance, ~\cite{BangOMPB20} partitions the hardware resources into virtual domains and 
allocates them to separate 
index instances to isolate interfering queries.
~\cite{BangOMPB20} does not partition the index. Rather, it identifies the optimal virtual domain size to schedule queries over pre-partitioned indexes. Notably, both of these works allow any core on the local NUMA node to schedule queries. Unlike \sys{}, they overlook the benefits of scheduling queries that access the same memory pages in nearby cores, which can result in sub-optimal schedules for servers with high core counts. 









\section{Background}
\label{sec:bk}
\noindent\textbf{Problem Statement: Spatial Scheduling}. 
Spatial scheduling~\cite{GicevaARH14} 
decides which core in the underlying hardware gets to execute a query (core scheduling), and which NUMA node in the underlying hardware gets to store the data requested or generated by the scheduled query (data partitioning). 
This is orthogonal to traditional query scheduling that is {\bf temporal} in nature, i.e., when to schedule a query. The core idea behind spatial scheduling is to minimize communication distance between two cores to account for intra-socket NUMA heterogeneity. The aim is to co-schedule queries that access common memory pages to nearby cores, while 
scheduling interfering queries to distant cores. To account for  inter-socket NUMA heterogeneity, the goal is to 
distribute data to 
NUMA nodes to maximize local memory access. 

\noindent\textbf{Offline Reinforcement Learning}~\cite{orl-levine-tut}. An agent is provided with an offline dataset that consists of various state-action transitions. These transitions represent past experiences. They are collected by running a variety of heuristic-based policies in different contexts. The goal of offline RL is to learn a scheduling policy exclusively from this offline dataset, without any active interaction with the DBMS. This differs from  traditional RL, where the agent learns a new scheduling policy, and then deploys it in the real-world, i.e., the DBMS, for feedback. In traditional RL, the agent goes through a continuous trial-and-error process, 
and progressive
learning 
until it reaches a stable state. This involves the agent taking suboptimal decisions over a long period that may degrade the DBMS performance in an unpredictable manner. 
In offline RL, there is no feedback loop between the agent and the DBMS. The agent is constrained to learn only from the scheduling policies in the provided offline dataset. Hence, in the learning phase, \sys{'s} learned agent does not interfere with the DBMS. This allows data-driven training of \sys{}'s learned agent. The offline RL learning process differs from  traditional RL. However, the agent's goal remains the same, i.e., to maximize a reward function, e.g., query throughput.  

\noindent\textbf{Hardware Profiling}. 
Processors have dedicated hardware blocks throughout the chip, termed Performance Monitoring Unit (PMU, for short) that tracks the hardware performance statistics~\cite{pmuIntel,pmuAMD,pmuARM}. Almost all hardware components, e.g., the cores, memory controllers, interconnects, PCIe lanes of the processor are equipped with their own PMU blocks. At the core of these PMU blocks, there are multiple counters (1 to 8 depending on the processor and the location of the PMU block) paired with a control register. Also, each PMU provides a fixed list of Performance Monitoring Events. Based on the location of the PMU (inside or outside a logical core), these events are classified into core and uncore (off-core) events. Some of the counters 
collect architectural monitoring events, i.e., events that behave consistently across hardware micro-architectures, e.g., the number of executed instructions, the number of LLC misses, the number of core cycles. The remaining counters 
monitor a single or multiple
Performance Monitoring Events through the control registers. \sys{} uses these hardware statistics
as they mimic
the data distribution and capture the query workload. For example, a key range  that is being queried extensively would result in a lot of cache and memory accesses. The number of cache accesses may further increase, if the data distribution within the key range is particularly dense. In the same spirit, a lookup query yields less cache and memory accesses compared to a range scan. 


\begin{figure*}[t]
    \captionsetup{belowskip=-10pt}
    \captionsetup{aboveskip=-4pt}
    \centering
    \includegraphics[width=\textwidth]{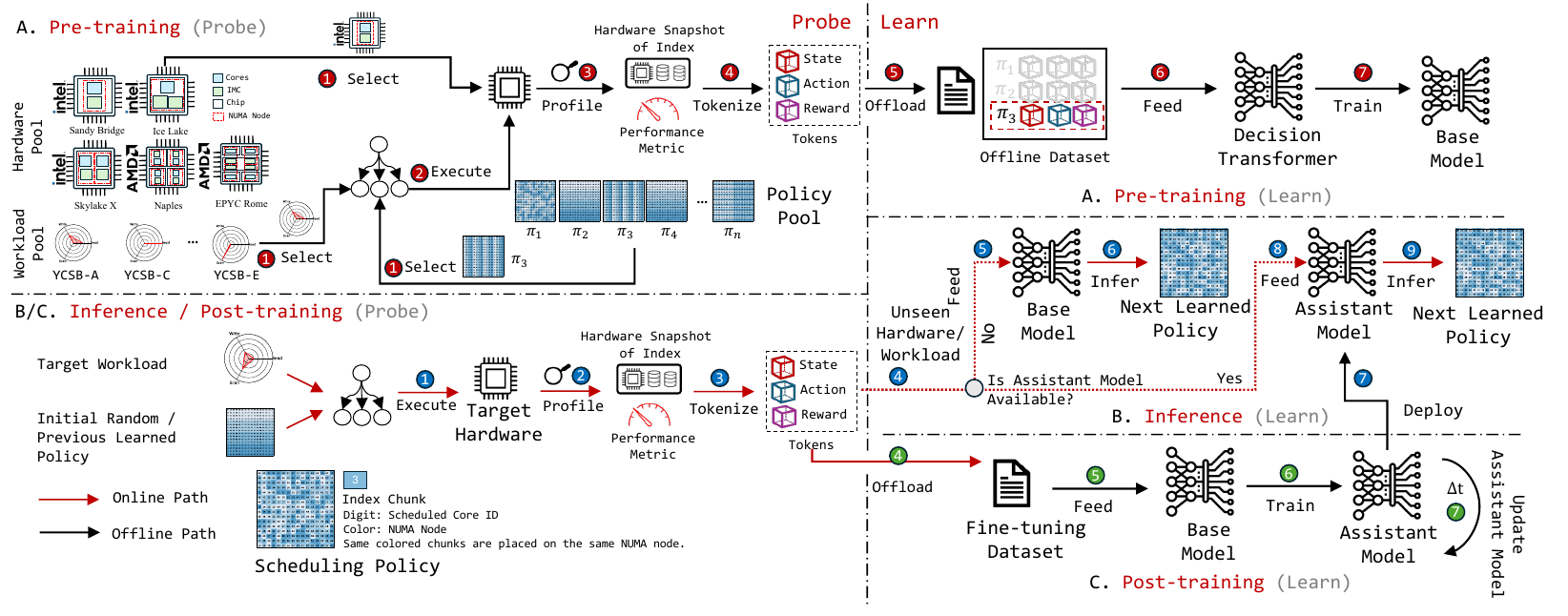}
    \caption{ \sys{} architecture.}
    \label{fig:arch}
\end{figure*}

\section{Overview of \sys{}}
\label{sec:overview}
Figure ~\ref{fig:arch} gives the \sys{} architecture. \sys{} follows the {\em Probe and Learn (PoLe)} technique~\cite{aidm25_ntp},
i.e., it probes the hardware during query execution to observe the hardware PMU statistics under a given scheduling policy, and gains insights from multiple 
hardware-DBMS kernel interactions to learn a better scheduling policy in an automated manner. 
The complete pipeline of \sys{} consists of three stages: pre-training, inference, and post-training.

\vspace{4pt}
\noindent\textbf{A. Pre-training.} During the pre-training phase, \sys{} curates a hardware pool comprising various NUMA and Chiplet servers from different CPU vendors, each with distinct NUMA configurations. In parallel, \sys{} constructs a workload pool comprising diverse workloads and a policy pool comprising different scheduling policies. Then,\!\ciro{1}\!\sys{} samples a specific query workload (e.g., YCSB-A) along with a scheduling policy (e.g., $\pi_3$) from the respective pools, and\!\ciro{2}\!executes the selected workload under the chosen policy on a chosen server (e.g., a 4-Socket Intel Sandy Bridge). During query execution,\!\ciro{3}\!\sys{} periodically \textit{probes} the hardware PMUs, and generates a \verb|Hardware Snapshot| of the index under the current scheduling policy. The \verb|Hardware Snapshot| consists of hardware statistics from various hardware parts, i.e., CPU cores, memory controllers, and off-chip interconnect networks. 
These hardware snapshots reflect the main-memory index state under the current scheduling policy from the perspective of hardware performance counters.\!\ciro{4}\!\sys{} tokenizes the \verb|Hardware Snapshot|, the current scheduling policy and the performance metric, generating state, action and reward tokens, respectively.\!\ciro{5}\!Then, these tokens are  periodically offloaded to an offline dataset. Once the offline dataset covers all the workloads, policies, and hardware from the respective pools,\!\ciro{6}-\!\ciro{7}\!\sys{} trains a Decision Transformer~\cite{ChenLRLGLASM21} (DT, for short) on the ``offline'' dataset in a supervised manner using offline RL. Finally, the pre-training stage yields a ``base'' learned model. The whole stage is conducted fully offline. 

\vspace{4pt}
\noindent\textbf{B. Inference.} The inference phase represents the online execution path of the main-memory index. As the 
index executes the target query workload on the target hardware, \sys{} actively infers new scheduling policies in response to any changes in the query workload. 
\!\ciroo{1}\!Initially, at 
Time
$t_0$, \sys{} bootstraps the main-memory index with a random scheduling policy, and executes the target workload under the chosen policy on the given target hardware.\!\ciroo{2}\!In parallel, \sys{} continuously monitors the hardware PMU statistics and\!\ciroo{3}\!tokenizes them in real time.\!\ciroo{5}\!These tokens are 
fed to the ``base'' model from the pre-training stage \!\ciroo{6}\!to infer the initial learned policy. \sys{} enforces the new learned scheduling policy instantly, by updating the mapping between each index slice and the CPU cores.\!\ciroo{4}\!-\!\ciroo{6}\!As the query workload evolves, the cycle continues. \sys{} tokenizes the hardware PMU statistics under the current scheduling policy and the target workload. Then, they are 
fed to the ``base'' model to infer subsequent scheduling policies. 

\vspace{4pt}
\noindent\textbf{C. Post-training Stage.} 
\!\ciroo{1}During query execution on the target hardware and the target workloads in the inference stage, 
\!\ciroo{2}\!-\!\cirooo{4}\!\sys{} periodically offloads the tokens of the target hardware and the target workloads to a dedicated ``fine-tuning'' dataset. Unlike the ``offline'' dataset used during pre-training, the fine-tuning dataset contains tokens that correspond to the target hardware, workloads, and the learned scheduling policies.\!\cirooo{5}\!\!-\cirooo{6}\!\sys{} trains the base model on the ``fine-tuning'' dataset using offline RL to generate the ``assistant'' model.\!\cirooo{7}\!As the query workload evolves, the ``fine-tuning'' dataset grows. \sys{}  updates the assistant model by incrementally training it on the newly collected tokens. Each iteration of the post-training stage yields an assistant model and is conducted fully offline. Once  post-training  completes,\!\ciroo{7}\!\!-\ciroo{9}\!\sys{} replaces the base model with the updated one, ensuring that \sys{}  uses the most recent assistant model during inference.

\vspace{4pt}
\noindent\textbf{Why Choose PMU and Offline RL as the Building Blocks of \sys{}?} Low-level statistics from the Hardware PMU and Offline RL are the core building blocks of \sys{}. A key challenge in basing \sys{}'s scheduling policy on PMUs alone is that the statistics collected from the PMUs can be non-deterministic, i.e., different runs can generate different values. However, the relative trends in these statistics remain consistent. Also, modern processors provide a large number of hardware events to profile. Hence, it becomes increasingly difficult to handpick the important hardware statistics that fit across different NUMA and Chiplet architectures and query workloads. 
This also makes it difficult to design heuristic-based schedules using PMU statistics. 
Similarly, a key challenge in employing offline RL in \sys{} is: How to compile an offline dataset where the state-action transitions can abstract diverse CPU vendors, architectures, query workloads, and data distributions under a generalized framework? This is necessary so that the agent learns to generalize across these different settings. Both PMU and offline RL work in cohesion to address these challenges. The hardware statistics of PMUs can abstract across these diverse settings to generate high-quality data, and prepare an offline dataset. Also, the ML techniques used in Offline RL are well suited to handle the possible inconsistencies in PMU data as they focus on modeling the wider trends present in the data. Together, they guide the scheduling process in \sys{}. 

\section{Learning Scheduling Decisions in \sys{}}
First, we discuss how \sys{} formalizes the Next Token Prediction (NTP) task of query scheduling over a main-memory index as a Reinforcement Learning (RL) task. 
Then, we present the design of the Decision Transformer (DT) model and discuss the token generation process that serves as the input to the DT. Finally, we explain \sys{'s} training and inference stages.

\subsection{NTP and Spatial Scheduling}
\label{sec:ntp_sched}
Next Token Prediction (NTP) is at the core of modern LLMs. By learning to predict the next token, LLMs build a deep understanding of the world, and can exhibit emergent behaviors. One important reason for the success of NTP can be attributed to the language tokens, i.e., words, sub-words, or letters. These language tokens are syntactically regular (e.g., a verb follows a subject), and the context is inherently baked into the tokens (e.g., `AMD Milan is a chiplet processor with 2 (1) sockets' conveys the characteristics of the Milan processors, i.e., it is from AMD, follows a Chiplet architecture and has a maximum of 2 sockets). Translating NTP to the scheduling task is analogous to predicting the next ``Core ID'' in a sequence. In this context, the Core IDs act as the tokens and the sequence corresponds to the scheduling policy. The
length of the scheduling policy remains fixed, equal to the number of logical partitions of the index. However, unlike language tokens, the Core IDs lack syntactic regularity and context. A Core ID itself (e.g., 1) does not capture the underlying hardware characteristics. Neither does it capture its impact on index performance, which is the ultimate goal of the scheduling task. Hence, to adopt the paradigm of NTP for scheduling task, these Core IDs need additional context. This aligns well with the RL paradigm, as RL inherently provides contexts to actions, i.e., the Core ID tokens in this context, through observed rewards and state transitions. 

\subsection{RL Formulation of Spatial Scheduling}
\begin{figure}[tbp]
    \captionsetup{belowskip=-20pt}
    \captionsetup{aboveskip=-0mm}
    \centering
    \includegraphics[width=\columnwidth]{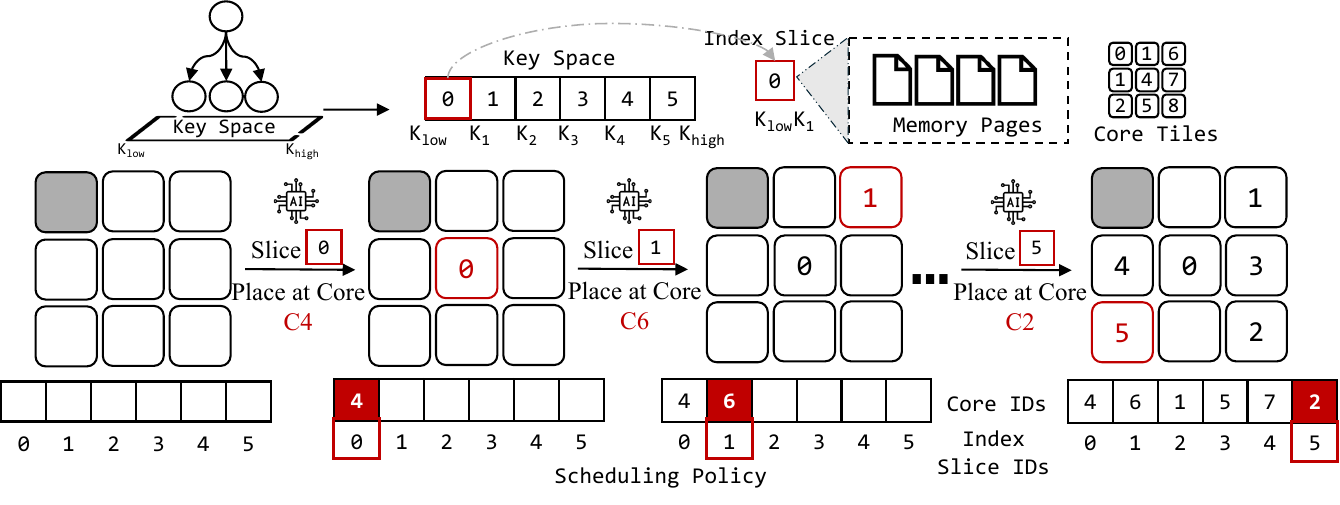}
    \caption{RL problem formulation of spatial query scheduling over a main-memory index. 
    }
    \label{fig:rl_pro}
\end{figure}

We take inspiration from the chip placement problem~\cite{MirhoseiniGYJSW21,LaiLTWH023}, and cast the NTP task of spatial query scheduling over a main-memory index as an RL problem. \sys{'s} RL agent   chooses each index slice one at a time, and sequentially schedules them on one of the worker cores. To maximize data locality, we assume there exists an inherent affinity between the core where the index slice is scheduled and the NUMA node where the index slice is placed, in terms of data placement. When an index slice is scheduled on a certain core, the index nodes within the slice are placed on the DIMM chip attached to the core's local NUMA node. Refer to Figure~\ref{fig:rl_pro}. At Time $t_0$, 
no 
index slices are placed on the hardware. At Time $t_1$, the agent predicts the next core, C4, to place the first index slice. At Time $t_2$, the agent predicts the next core, C6, to place the second index slice. This process continues until at Time $t_T$, \sys{} places the last index slice in Core C2. The non-worker cores (marked dark gray) are unavailable for the agent to choose from. 

\vspace{4pt}
\noindent\textbf{Agent-Environment Interface}. Formulating an RL problem is based on four elements denoted by the tuple $\langle\mathcal{S, A, P, R}\rangle$. Let $s_t, a_t$, and $r_t = R(s_t, a_t)$ be the respective state, action, and reward at Time Step $t$. Then, $\langle s_0, a_0, r_0, s_1, a_1, r_1, \ldots, s_t, a_t, r_t\rangle$ denotes a spatial scheduling policy of a main-memory index. We give an overview of these elements from the perspective of \sys{} as follows. 

\vspace{4pt}
\noindent1. \textbf{State ($\mathcal{S}$)}: Given a main-memory index under a certain scheduling policy, \sys{} uses the hardware PMU statistics (e.g., instruction count, LLC misses) observed at different cores, the traffic in  memory channels and interconnect links, 
and the current placement of the index slices to represent the index and the hardware state. 

\vspace{2pt}
\noindent2. \textbf{Action ($\mathcal{A}$)}: Given the next index slice to schedule, the action space of \sys{} consists of worker cores eligible for selection, i.e., prediction, without violating any constraints.

\vspace{2pt}
\noindent3. \textbf{State Transition ($\mathcal{P}$)}: Given the current state and an action, state transition captures how to transition to a new state and reflects the aftermath of the latest placement decision. When the last index slice is placed, the agent reaches a terminal state. 

\vspace{2pt}
\noindent4. \textbf{Reward ($\mathcal{R}$)}: As an index slice is placed into a CPU core, $\mathcal{R}$ reflects how this  decision improves  index query throughput.  

\subsection{Model Architecture}
\label{sec:dt}
After formulating the query scheduling task as an RL problem, we need to select the appropriate learning paradigm to solve it. Traditional learned DBMS components adopt \textit{online} RL, where the agent operates in the critical path of query execution. 
It continuously interacts with the DBMS via a feedback loop to balance exploration and exploitation that can lead to sub-optimal query performance. This also requires the agent to be retrained from scratch in heterogeneous cloud environments where the hardware and workload are not known beforehand. To isolate the agent's learning process from the critical path of query execution, \sys{} adopts Offline RL, precisely a Decision Transformer~\cite{ChenLRLGLASM21}. \sys{} constrains its RL agent to learn the scheduling policy from a pre-collected dataset without interrupting DBMS execution, enabling \sys{} to leverage its pre-trained base model to immediately generate high-quality schedules \textit{within minutes}, even in previously unseen environments.

\vspace{4pt}
\noindent \textbf{Decision Transformer}. 
\sys{} adopts the Decision Transformer~\cite{ChenLRLGLASM21} (DT) as the model architecture for its RL agent. A DT is an auto-regressive model that predicts a future action based on the past actions, observations, and a desired reward. 
DT abstracts RL as a sequence modeling problem, and learns the scheduling policy in a supervised manner by conditioning on a desired query throughput objective. DT uses  offline RL, i.e., during the pre-training and the post-training stages, it is trained  on an offline and fine-tuning dataset without any interaction with the index. 

Let $s_t, a_t$ and $\hat{r}_t$ be the respective state, action and desired reward at Time Step $t$. Let $\bm{{s}_{<t}}, \bm{{a}_{<t}}$ and $\bm{\hat{r}_{<t}}$ be the respective state, action, and desired reward in the time steps preceding $t$. The DT estimates the probability distribution of the next action $a_t$ at Time Step $t$, i.e., $\Prob(a_t|\bm{a_{<t}},\bm{s_{\leq t}, \bm{\hat{r}_{\leq t}}})$. Following the seminal work in~\cite{ChenLRLGLASM21}, the GPT-1 model~\cite{gpt18} serves as the backbone for \sys{}'s DT. GPT is a sequence-to-sequence model that lies at the core of modern LLMs. GPT replaces the softmax layer of the original Transformer architecture~\cite{VaswaniSPUJGKP17} with a causal self-attention mask so that it can be trained in an auto-regressive manner by only attending over the previously seen inputs, in contrast to attending over all the inputs. 

In LLMs, the input to GPT is a sequence of $T$ tokens, where each token may correspond to a character, sub-word, or word. $T$ refers to the input sequence length. In contrast, in DT, the GPT is fed a total of 3T+1 tokens for a single scheduling policy, with T tokens each for the state, action, and reward modalities, and an additional meta-token for representing the system-wide view of the hardware. These tokens are called \textit{State}, \textit{Action}, \textit{Reward-To-Go} (RTG, for short) and \textit{Meta} tokens, respectively. $T$ refers to the number of index slices to schedule in this context. First, the State token is passed through a convolutional encoder consisting of three convolutional layers, followed by a fully connected layer to obtain the State-token embedding. The Action and RTG tokens are passed through separate embedding layers to obtain the respective embeddings. Once  embeddings are learned, they are passed to the GPT architecture that predicts the next action through auto-regression. 
To formulate spatial scheduling using NTP (\S\ref{sec:ntp_sched}), the Action tokens are analogous to the tokens to be predicted. The State, RTG, and  Meta tokens provide the Action tokens with the necessary context for effective prediction. The final output of \sys{'s} DT is a sequence of actions, i.e., Core IDs, that correspond to the $i$-th index slice. 

\vspace{4pt}
\noindent\textbf{Why Decision Transformer (DT)}? 
Mainly, there exists three classes of Offline RL algorithms; Q-Learning (e.g., ~\cite{KumarZTL20,FujimotoMP19,KostrikovNL22}), Imitation Learning (e.g., ~\cite{FujimotoG21}), and Transformer-based methods (e.g.,~\cite{ChenLRLGLASM21,ZhengZG22}). Each class has its own pros and cons. Q-Learning suffers from instability and needs extensive hyperparameter tuning, while  Imitation-Learning requires the policies in the offline dataset to be of very high quality. This makes these two approaches less appealing.
\sys{} 
relieves the DBMS kernel from hand-tuning the scheduling policies through a learned agent. Thus, spending considerable efforts in tuning the hyper-parameters of the learned agent is rather contradictory. Moreover, it is quite natural for the offline dataset to include sub-optimal policies, especially from  the DBMS scheduling perspective. The reason is that no single scheduling policy can dominate across all query workloads, data distributions, and NUMA configurations (See \S\ref{sec:eval}). As the query workload, data distribution, or the hardware changes, the same scheduling policy is likely to yield sub-optimal performance. Transformer-based methods model very large sequences and are highly scalable. They provide stable training, and greater generalizability that fits 
\sys{}. 
For spatial scheduling using RL, the number of index slices refers to  the sequence length that is set to 256 in \sys{}. Moreover, \sys{} uses a large number of hardware counter statistics, typically in the range of [15, 39], and strives for adaptivity across different CPU vendors, NUMA architectures, query workload, and data distribution. 
Hence, 
DT is particularly appealing to \sys{}. 

\begin{figure*}[t]
    \captionsetup{belowskip=-10pt}
    \captionsetup{aboveskip=0pt}
    \centering
    \includegraphics[width=\textwidth]{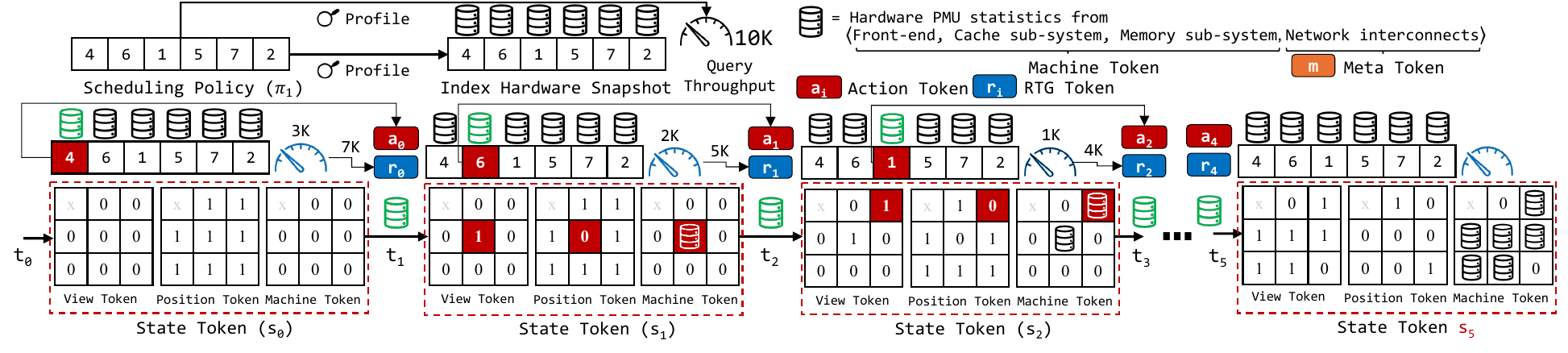}
    \caption{Tokenization in \sys{}.}
    \label{fig:tokens}
\end{figure*}

\subsection{DT Features and Tokens}
\label{sec:token}

\noindent\textbf{Feature Representation in \sys{}}. 
Recall from Section~\ref{sec:overview} that \sys{} periodically probes the hardware PMU registers during query execution to collect hardware PMU statistics across all three stages of the pipeline. 
\sys{} chooses these hardware PMU statistics as the feature representation for the DT over any software-level bookkeeping, e.g.,~\cite{AnneserKZ0K22, StoicaA13}. DBMS kernels that rely on software-level bookkeeping for optimization, the bookkeeping process itself can become a performance bottleneck as system complexity or workload intensity increases. In contrast, hardware performance statistics offer fine-grained, low-level measurements collected directly at the hardware level, providing the highest possible granularity with significantly lower collection overhead. 

To enable abstraction across diverse hardware configurations, \sys{} profiles the following \textit{four} hardware components: the front-end, the cache sub-system, the memory sub-system, and the network interconnects for each index slice. \sys{} profiles a maximum of 39 hardware counters from these components, including 15 core counters and 24 off-core counters (for Intel servers only). These counters are selected based on two criteria: their availability across different CPUs, and their ability to capture index performance. The selection is consistent with prior micro-architectural analyses~\cite{vtune,AilamakiDHW99}. 
The hardware PMU statistics from the front-end include the number of instructions executed, L1-I (L1-Instruction) cache misses, and branch mispredictions per 1k instructions. The cache sub-system statistics include the number of L1D (L1-Data), LLC (Last Level Cache), DTLB (Data Translation Lookaside Buffer), and LLC-write misses per 1k instructions. The memory sub-system statistics include the number of memory accesses, retired load instructions serviced from local and remote DRAM, and memory-write misses per 1k instructions. The network interconnect statistics include the system-wide bandwidth usage of each memory channel of the NUMA nodes, i.e, Integrated Memory Controllers, and the amount of incoming and outgoing traffic in the off-chip interconnects. 
For each index slice, these corresponding hardware PMU statistics from four components are combined to generate a \verb|Hardware Snapshot| of the main-memory index. Refer to Figure~\ref{fig:tokens}. Each index slice is uniquely associated with its own set of hardware PMU statistics. Note that the hardware PMU may use the same or different event names to profile a particular hardware event. Hence, a mapping is required to align platform-specific hardware events to the features stated above.

\vspace{4pt}
\noindent\textbf{Tokenization}. As discussed in Section~\ref{sec:dt}, in each time step $t$, the DT is fed three separate tokens: a State token $s_t$, an Action token $a_t$, and an RTG token $\hat{r}_t$ along with a separate Meta token $m$. Refer to Figure~\ref{fig:tokens} for an illustration of the tokenization process that builds on the example in Figure~\ref{fig:rl_pro}. The hardware PMU statistics from the front-end, cache-subsystem, and the memory subsystem generate the State token, while the hardware PMU statistics from the network interconnects generate the Meta token. The corresponding spatial scheduling policy $\pi_1$, and the performance metric, i.e., the query throughput, produce the Action and RTG tokens, respectively. 

\vspace{4pt}
\noindent\textbf{1. State Token}: Given the latest (partial) spatial query scheduling policy, State tokens represent the current state of the hardware. \sys{} represents the hardware as a $C_{m_i\times m_j}$ core tile, where $|C_{m_i\times m_j}|$ refers to the total cores in the system. \sys{} fuses three distinct tokens: a {\em View} token, a {\em Position} token and a {\em Machine} token to generate the State token. Each token reflects the hardware state from a different perspective. Only the worker cores in the hardware are eligible for scheduling. Cores reserved for routing or collecting hardware traces do not participate in the scheduling process. 
Next, we discuss each of these three tokens. 

\noindent\textbf{1a. View Token} 
refers to the occupied core tiles in the hardware. Each entry in the View token indicates if the corresponding core has been scheduled to host any index slice or not. In Figure~\ref{fig:tokens}, at Time $t_0$, all  worker cores are empty as no scheduling decision has taken place yet. Thus, the View token consists of all 0s'. At Time $t_1$,  Policy $\pi_1$ chooses Core C4 to schedule the first index slice. Hence, the corresponding entry of Core C4 in the View token is set to 1. 
Note that multiple index slices can be scheduled on the same core allowing \sys{} to schedule larger indexes, when the number of index slices exceeds the number of available worker cores.

\noindent\textbf{1b. Position Token}
refers to the core tiles eligible for scheduling the next index slice. Each entry in the Position token with Value ``1" indicates that the corresponding core is available for scheduling, whereas an entry with Value ``0" indicates the corresponding core is not available. Non-worker cores are made unavailable for scheduling from the start (e.g., Core C0 in Figure~\ref{fig:tokens}). The Position token can impose additional scheduling constraints, e.g., ensuring each worker core gets at least $n$ index slices, or the next index slice is placed in the vicinity of the previous index slice in the hardware, etc. \sys{} imposes no such constraints during training, but enforces an $n$-slice-per-core constraint during inference. In Figure~\ref{fig:tokens}, $n=1$, i.e., at Time $t_1$, when Policy $\pi_1$ chooses Core C4 to schedule the first index slice, the corresponding core tile of Core C4 is set to ``0'', and it remains so for subsequent time steps.

\noindent\textbf{1c. Machine Token} captures the hardware state under a given scheduling policy from the perspective of the front-end, the cache and memory subsystems of the hardware. 
In Figure~\ref{fig:tokens}, the Machine token at Time $t_0$ is empty. At Time $t_1$, under Policy $\pi_1$, the first index slice is scheduled on Core C4. Thus, the Machine token is updated by aggregating the hardware snapshot of the first index slice with the corresponding entry of Core C4. When multiple index slices are scheduled on the same core, the criterion to aggregate the hardware snapshots for the particular index slices can vary. \sys{} has addition as the aggregation criterion. Note that only the hardware PMU statistics from the front-end, cache subsystem, and the memory subsystems are used to generate the machine token. 

\vspace{4pt}
\noindent\textbf{2. Meta Token} captures the hardware state under a given scheduling policy from the perspective of the network interconnects of the hardware. Contrary to the \textit{Machine} token that represents the hardware state at the CPU core level, a meta token represents the hardware state at the system-wide level, e.g., the state of the Integrated Memory Controllers, the off chip interconnects, etc. Aside from the hardware statistics from the network interconnects, the Meta token includes  processor-level information of the hardware, e.g., the number of logical cores, the number of NUMA nodes, the number of sockets, and the CPU vendor as domain knowledge.

\vspace{4pt}
\noindent \textbf{3. Action Token} has a state space of all possible worker cores, (Cores C1-C8 in Figure~\ref{fig:tokens}). At Time $t_1$, as the  index slice is scheduled in Core C4, ``4'' is  the Action token for Time $t_1$. Similarly, Core ``6'' is the Action token for Time $t_2$, ``1'' for Time $t_3$, etc. 

\vspace{4pt}
\noindent\textbf{4. The Return-To-Go Token (RTG)} is calculated as the difference between the desired objective, i.e., query throughput and the rewards observed so far. \sys{} uses the index query throughput as the reward criterion for the RL agent. During training, the initial RTG Token at Time $t_0$ is set to 1.0$\times$ the query throughput that the scheduling policy achieves, e.g., 10K (cf. Figure~\ref{fig:tokens}). During inference, the initial RTG token at $t_0$ is set to the maximum query throughput the index targets to achieve. As index slices are scheduled sequentially, the RTG token is updated. At Time $t_1$, the agent places the first index slice on Core C4 yielding a 3K throughput. Thus, the RTG for $t_1$ is updated accordingly to 7K.

\subsection{Training}
Inspired by LLMs, \sys{} trains the DT in two stages: a pretraining stage referred to as generative pretraining, and a post-training stage referred to as fine-tuning. An offline dataset guides the pretraining stage while a fine-tuning dataset guides the post-training stage. During both stages, \sys{} employs offline RL, and follows the Teacher Forcing strategy~\cite{WilliamsZ89} to train the DT. 

\vspace{4pt}
\noindent\textbf{Offline Dataset Creation}. In the 
pre-training stage, \sys{} curates a hardware pool comprising a 1-NUMA-Per-Socket AMD Milan, a 4-NUMA-Per-Socket AMD Milan, a 4-Socket-8-NUMA Node Intel Skylake X, a 4-Socket Intel Sandy Bridge, and an NVIDIA H200 Grace Hopper server to create the offline dataset that consists of the YCSB benchmark. The query workload comprises 50\% Read-50\% Write, 100\% Read, 100\% Scan, 20\% Read-30\% Scan-50\% Insert, 25\% Read-50\% Scan-25\% Insert workloads. The scheduling policies cover the heuristics: grouped (14.63\%), mixed (12.75\%), spread (1.76\%), and random strategies (71.03\%). In the generative pre-training, \sys{} selects a specific dataset, e.g., YCSB, along with a query workload, e.g., a Zipfian 100\% Read workload. \sys{} runs the selected query workload on one of the servers, e.g., NVIDIA H200 Grace Hopper. As the query workload executes, \sys{} profiles the hardware to generate the index hardware snapshot, and tokenizes them to generate the State ($s_t$), Action ($a_t$), RTG ($\hat{r}_t$) and Meta tokens ($m$) (as in Section~\ref{sec:token}). These tokens are  offloaded to the offline dataset, where each sample 
is of the form: $\langle \hat{r}_0, s_0, a_0, \hat{r}_1, s_1, a_1, \ldots, \hat{r}_{T-1}, s_{T-1}, a_{T-1}\rangle$. $T$ denotes the number of index slices. 

\vspace{4pt}
\noindent\textbf{Fine-tuning Dataset Creation}. During 
post-training,
\sys{} constructs a fine-tuning dataset comprising tokens from various learned scheduling policies evaluated on the target hardware for the target workloads. Unlike the offline dataset, the samples from the fine-tuning dataset belong to a distinct hardware, e.g., an Intel Skylake X machine. As the query workload evolves, \sys{} continuously incorporates tokens from different learned scheduling policies under different target query workloads in the fine-tuning dataset. The samples in the fine-tuning dataset maintain the same format as those in the offline dataset.

\begin{figure}[t]
    \captionsetup{belowskip=-10pt}
    \centering
    \includegraphics[width=0.95\columnwidth]{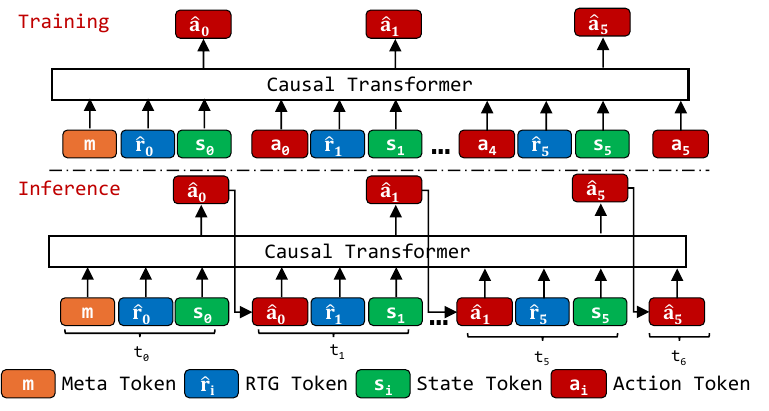}
    \caption{Training and Inference stages in \sys{}.}
    \label{fig:pmoss_learned}
\end{figure}

\vspace{4pt}
\noindent\textbf{\sys{'s} Training Loop}
is consistent across both training stages. The difference lies in the datasets used. In  pre-training, the DT is trained on an offline dataset. However, in post-training,  the DT is trained on a fine-tuning dataset that comprises learned scheduling policies on  target hardware and workloads. Figure~\ref{fig:pmoss_learned} shows \sys{'s} DT  training. In pre-training, \sys{} begins with an untrained DT to generate the ``base'' model. Post-training resumes from the base model, and generates an ``assistant'' model.

At Time $t$, DT processes the Meta token $m$ along with the most recent $K$ triplet tokens: the State token, the Action token, and the RTG token to predict the next Action token $\hat{a}_{t}$, i.e., the Core ID for the $t$-th index slice. $K$ refers to the context length of DT that is set to 256; the number of index slices in \sys{}. \sys{} trains the DT in a supervised manner 
following the Teacher Forcing strategy~\cite{WilliamsZ89}. 
The objective is to align the predicted Action tokens with the Action tokens present in the offline (fine-tuning) dataset referred to as the ground-truth Action tokens, using the standard cross-entropy loss as follows. Let $\hat{a}_t$ and $a_t$ be the predicted and ground truth Action token at Time $t$. Let $\bm{{s}_{<t}}, \bm{{a}_{<t}}$ and $\bm{\hat{r}_{<t}}$ be the respective State, Action and RTG tokens prior to Time $t$. Let $m$ be the Meta token. Then, 

$$\mathcal{L}_{\text{total}} = - \frac{1}{K} \sum_{t=0}^{K-1} \log \Prob \left( \hat{a}_t = a_t \mid \bm{s}_{\leq t}, \bm{a}_{<t}, \bm{\hat{r}}_{\leq t}, \bm{m} \right)$$

This supervised approach makes the training of the RL agent stable and scalable, in contrast to the traditional Bellman's Equation-based RL objective that can be brittle depending on the hyper-parameter tuning~\cite{ChenLRLGLASM21}. Both the pre- and post-training stages are conducted offline, ensuring that the learning process of \sys{} is decoupled from the index operations. \sys{} executes the 
pre-training stage once, while it executes the fine-tuning stage at regular intervals as the fine-tuning dataset grows over time.

\subsection{Inference} 
Figure~\ref{fig:pmoss_learned} illustrates \sys{'s} inference stage. If an assistant model from the post-training stage is unavailable, \sys{} defaults to using the base model  from the pre-training stage. Else, \sys{} uses the most recent assistant model for inference. 
At Time $t_0$, the trained DT is fed the Meta token ($m$), the initial State ($s_0$), and the RTG token ($\hat{r}_0$) that in turn generates the next Action token ($\hat{a}_0$), i.e., the Core Id to schedule the first index slice. \sys{} observes the recent Hardware Snapshots of the main memory index to obtain the consequent state ($s_1$) and the RTG token ($\hat{r}_1$), following the predicted Action ($\hat{a}_0$). At Time $t_1$, both the State and the RTG tokens, along with the predicted Action token are fed back into DT to predict the next Action ($\hat{a}_1$), i.e., the Core Id to schedule next index slice. This process repeats until the scheduling policy, i.e., the Core Ids for all the index slices are rolled out. $\langle\hat{a}_0,\hat{a}_1, \ldots, \hat{a}_{T-1}\rangle$ denotes a complete learned scheduling policy inferred by \sys{}. 

\section{Runtime System}
\label{sec:pmoss_sys}
\begin{figure}[t]
    \captionsetup{belowskip=-15pt}
    \captionsetup{aboveskip=0pt}
    \centering
    \includegraphics[width=1.0\columnwidth]{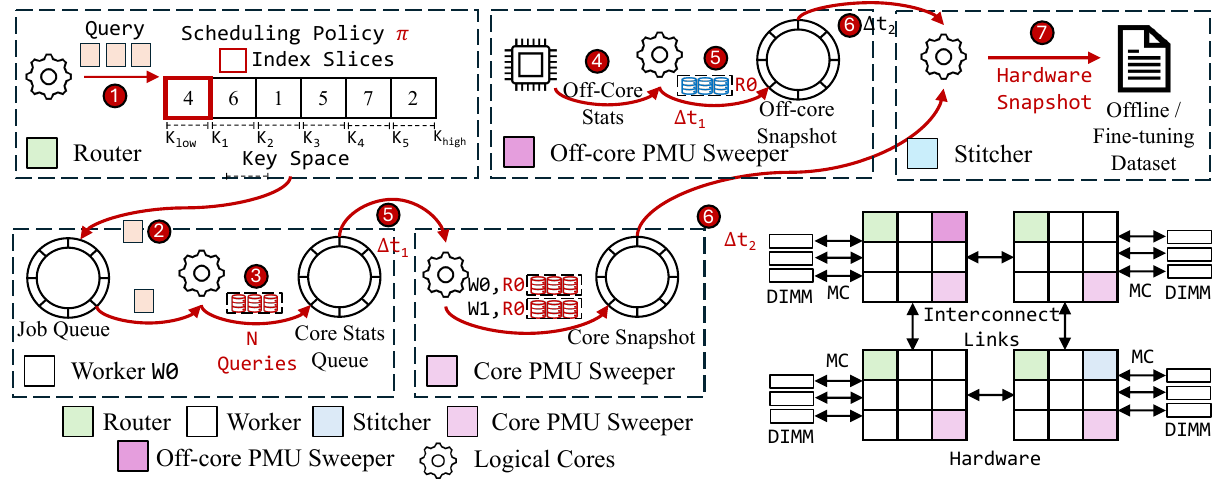}
    \caption{Life of a query inside \sys{}'s runtime system.
    }
    \label{fig:pmoss_system}
\end{figure}

\sys{} builds on top of a main-memory B$^+$-Tree implementation~\cite{WangPLLZKA18,btreecode} that follows  optimistic lock coupling (OLC). The B$^+$-Tree leaf node can hold a maximum of 255 entries. Upon start, \sys{} creates as many threads as the number of cores in the underlying hardware, and binds each thread to a core. Due to this, we use threads and cores interchangeably. In \sys{}, each core performs a diverse set of tasks ranging from routing, executing queries, to capturing hardware snapshots. Figure~\ref{fig:pmoss_system} illustrates the life of a query inside the runtime system of \sys{}. 

\vspace{4pt}
\noindent\textbf{Routing}. \sys{} maintains a number of router cores distributed equally among the NUMA nodes. Even though, in our experiments, \sys{} assigns a single core from each NUMA node of the target machine as a routing core, \sys{} can allocate additional routing cores as needed for servers with high core counts.\!\ciro{1}\!The router cores route the incoming queries to the intended worker cores for execution. Each router core in \sys{} maintains a copy of the scheduling policy. It evaluates the predicate of an incoming query to identify the index \textit{slice}, i.e., the key range it belongs to, and uses the stored mapping to route the query to its destination core. If there are multiple cores to choose from (e.g., a range scan may overlap multiple key ranges), the query's destination core is chosen randomly to balance the load. 

\vspace{4pt}
\noindent\textbf{Query Execution}. Most cores in \sys{} are worker cores that execute queries involving index traversal, and then produce results. Each worker core maintains two queues: A \textit{job} and a \textit{core stats} queues. A job queue stores the queries, while a core stats queue stores the hardware traces of an already executed query.\ciro{2}The router cores enqueue queries into the job queue. A worker core dequeues one query at a time from the job queue and executes it. 

\vspace{4pt}
\noindent\textbf{Profiling Core PMU Statistics}. Before query execution, the worker core programs the corresponding \textit{core} PMU counters of its PMU block to profile the compute, cache, and memory characteristics of the query.\ciro{3}It profiles the corresponding front-end, the cache and memory subsystems of the hardware to generate the Machine tokens (\S\ref{sec:token}). Once the query execution completes, it stops profiling, and inserts the hardware trace of the query at the tail of the core stats queue. The profiling granularity can impact  system performance as it entails invoking kernel functions. In \sys{}, we set the granularity to 100, implying
that the hardware trace contains the cumulative core PMU statistics of 100 executed queries. To profile core PMU statistics, \sys{} integrates a C++ wrapper for Linux Perf Event API~\cite{perf}, and reads the hardware statistics off the PMU counters in real time. 

\vspace{4pt}
\noindent\textbf{Profiling Off-core PMU Statistics}. \sys{} maintains an off-core sweeper core to collect the hardware traces left by a query in the \textit{off-core} components, e.g., the network interconnects. The off-core components do not distinguish between the data requests they receive from the different cores, and hence, we cannot isolate these statistics on a per core (query) basis. Thus, we collect these statistics at the NUMA node, or system levels.\ciro{4}In \sys{}, the off-core sweeper core  periodically profiles the memory controllers and interconnect PMUs to generate the Meta token (cf. \S~\ref{sec:token}). To profile off-core PMU statistics, \sys{} integrates Intel PCM~\cite{pcm}.

\vspace{4pt}
\noindent\textbf{Sweeping Round}. Each sweeping round, e.g., R0 includes sweeping the core HW statistics from each worker core, and sweeping the off-core HW statistics from the off-core PMU sweeper. \sys{} maintains a 
core 
PMU sweeper core in each NUMA node.\ciro{5}It sweeps the core statistics from the local worker cores, and creates a (partial) core snapshot of the corresponding NUMA node in the hardware. Similarly, the off-core sweeper core creates the off-core snapshot of the hardware. Both core and off-core snapshots  are stored in the respective sweeper core's snapshot queue. To accelerate the sweeping process, the sweeper cores use SIMD load and SIMD arithmetic instructions (add, multiply, divide) to sweep the core statistics from the local worker cores and generate the Hardware Snapshot of the main-memory index, respectively. During the stitching stage, \sys{} periodically offloads the Hardware Snapshot of the main-memory index by leveraging the SIMD store instruction.

\vspace{4pt}
\noindent\textbf{Stitching}.\ciro{6}One \sys{} core periodically collects the core and off-core snapshots from different sweeping rounds, and stitches them together to generate a complete \verb|Hardware Snapshot| of the 
index under the current scheduling policy. Occasionally, the same core serves as the enforcer to enforce the latest learned scheduling policy. This involves migrating each index slice to its designated NUMA node, and updating the mapping policy stored in each router core. Recall that, in \sys{}, each index slice logically maps to a key range. For a given index slice, the migration is 
performed 
by issuing a 
migratory range scan over the entire key range.
A migratory range scan migrates the index nodes that it touches to its destination NUMA node. \sys{} uses the Linux \verb|migrate_pages| to migrate index nodes across NUMA nodes. On average, it takes 0.1 seconds to migrate an index slice, and up to 25 seconds to migrate a B$^+$-Tree with 1000M records using a single thread.





\begin{figure*}[t]
\captionsetup[subfigure]{aboveskip=1pt}
\captionsetup[subfigure]{belowskip=2pt}
    \centering
    \begin{subfigure}[t]{\columnwidth}
        \centering
        \includegraphics[width=0.95\columnwidth]{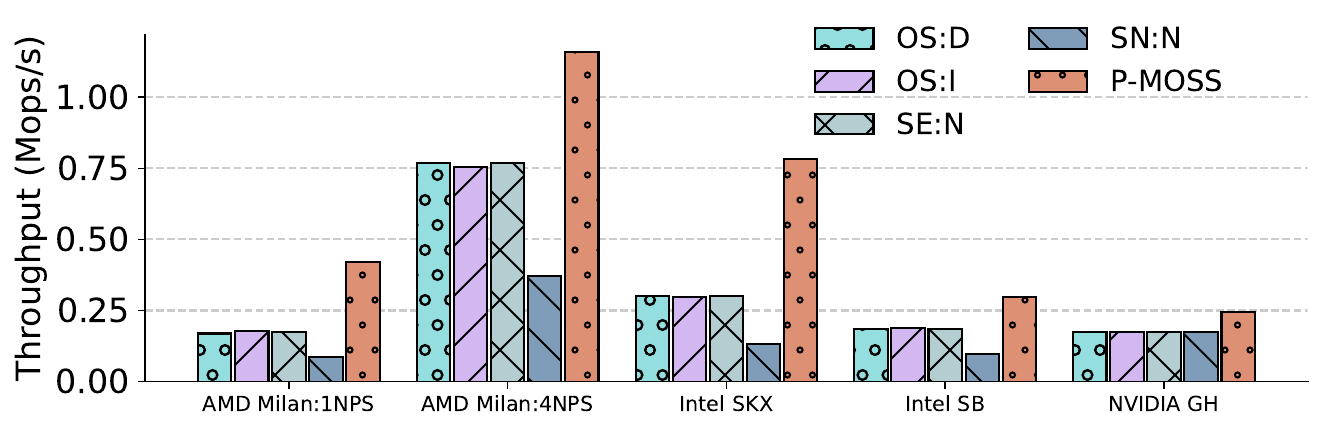}
        \caption{\sys{} vs OS baselines.}
        \label{fig:ycsba_linux}
    \end{subfigure}
    \begin{subfigure}[t]{0.95\columnwidth}
        \centering
        \includegraphics[width=\columnwidth]{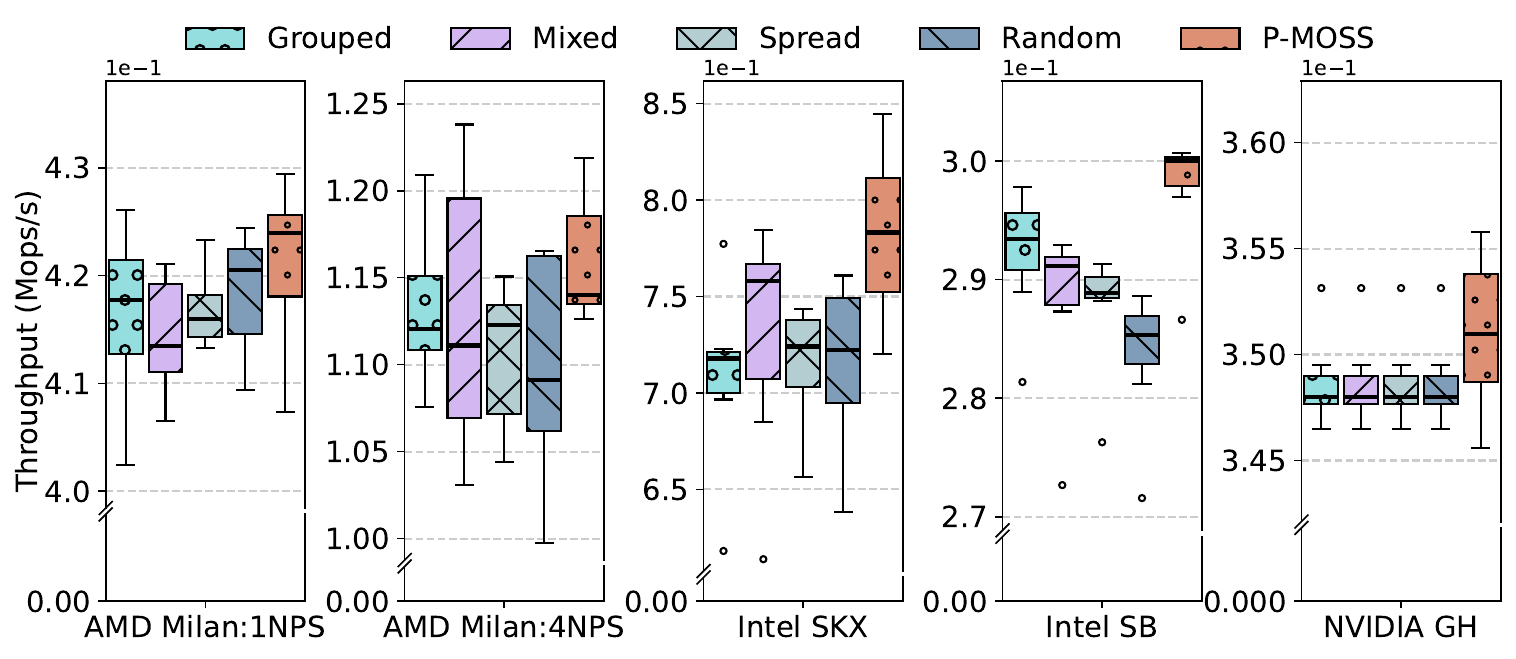}
        \caption{\sys{} vs SN:T baselines. }
        \label{fig:ycsba_learned}
    \end{subfigure}
\captionsetup{
belowskip=-10pt, aboveskip=-0.1pt}
\caption{\sys{} vs baselines on YCSB 
    50\% Read--50\% Write workload.}
\label{fig:ycsba}
\end{figure*}
\begin{figure*}[t]
\captionsetup[subfigure]{aboveskip=1pt}
\captionsetup[subfigure]{belowskip=2pt}
    \centering
    \begin{subfigure}[t]{0.95\columnwidth}
        \centering
        \includegraphics[width=\columnwidth]{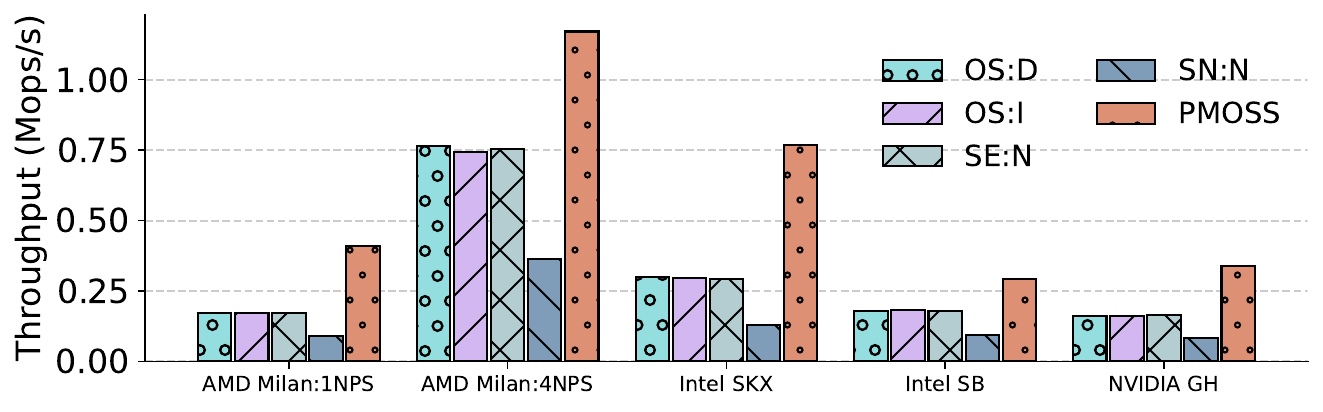}
        \caption{\sys{} vs OS baselines. 
        }
        \label{fig:ycsbc_linux}
    \end{subfigure}
    \begin{subfigure}[t]{0.95\columnwidth}
        \centering
        \includegraphics[width=\columnwidth]{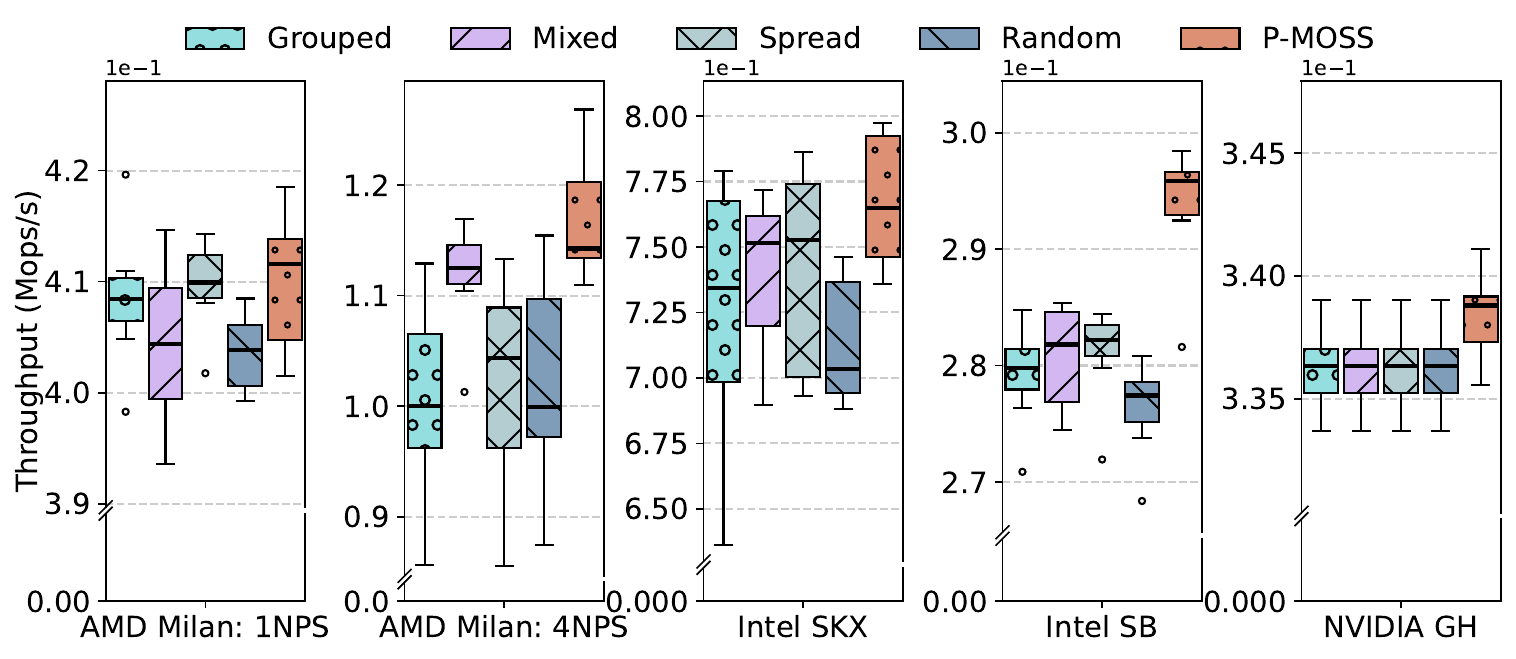}
        \caption{\sys{} vs SN:T baselines. 
        }
        \label{fig:ycsbc_learned}
    \end{subfigure}
\captionsetup{
belowskip=-10pt, 
aboveskip=-0.1pt
}
\caption{\sys{} vs baselines on YCSB Lookup workload.}
\label{fig:ycsbc}
\end{figure*}

\section{Evaluation} 
\label{sec:eval}
In this section, we study the performance of \sys{} in terms of query throughput against competing baselines. Specifically, we want to answer the following questions:
\begin{itemize}[leftmargin=*]
    \item How does \sys{} perform across different CPU vendors, e.g., the Intel, AMD, ARM and IBM processors?
    \item How does \sys{} perform across different NUMA architectures, e.g., Intel's cluster on die (COD) or the sub-NUMA clustering (SNC), and the Chiplet architectures, e.g., AMD's 1 NUMA per socket (1NPS), 4 NUMA per socket (4NPS) architecture?
    \item How does \sys{} perform for various query workloads? 
    \item How does \sys{} perform on unseen NUMA and Chiplet machines and query workloads?
    
    \item What is the overhead of collecting hardware profiles in \sys{} and how does SIMD benefit \sys{}? 

    \item How does the performance of  alternate RL techniques compare against the DT component of \sys{}?
    
    \item How does \sys{} perform under dynamic scenarios?
\end{itemize}

The next sections present the experimental settings and the results of our investigation to answer each of these questions.

\subsection{Experimental Settings}
\label{sec:exp_server}
\noindent\textbf{Testbed}. We evaluate \sys{} on a diverse set of NUMA machines ranging from different CPU vendors with different NUMA architectures. The details of the NUMA machines are as follows.

\vspace{2pt}
\noindent 1. \textbf{AMD Milan (1NPS)} is a dual socket 64 ($\times$2) core machine running Ubuntu 22.04 on a CloudLab~\cite{DuplyakinRMWDES19}  \verb|r6525| node. The processor is an AMD 7543 series, where  the BIOS setting for NPS (NUMA Per Socket) is set to default (=1). 

\vspace{2pt}
\noindent 2. \textbf{AMD Milan (4NPS)} is a dual socket 64 ($\times$2) core machine running Ubuntu 22.04 on a CloudLab~\cite{DuplyakinRMWDES19}  \verb|r6525| node. The processor is an AMD 7543 series, where the BIOS setting for NPS (NUMA Per Socket) is 4, i.e., each Socket is divided into 4 NUMA domains. Both AMD machines are equipped with 251GB of DDR4 RAM. 

\vspace{2pt}
\noindent 3. \textbf{Intel Skylake X} is a 4 Socket 92 ($\times$2) core machine running  Ubuntu 22.04 with 2.95TB of DDR4 RAM. Sub-NUMA Clustering is enabled. Thus, each socket is divided into 2 NUMA domains. 

\vspace{2pt}
\noindent 4. \textbf{Intel Sandy Bridge} is a 4 socket 32 ($\times$2) core machine running Ubuntu 22.04 on a CloudLab~\cite{DuplyakinRMWDES19} \verb|d820| node with 128GB DDR4 RAM. It has an Intel Xeon E5-4620 processor. It does not support Sub-NUMA clustering (SNC) or Cluster On Die (COD) technology. 

\vspace{2pt}
\noindent 5. \textbf{NVIDIA H200 Grace Hopper} is a single socket 72 ($\times$1) core machine running Ubuntu 22.04 on a CloudLab \verb|nvidiagh| node. It is based on ARM's Neoverse V2 architecture with 480GB DDR4 RAM.

\vspace{4pt}
\noindent \textbf{Baselines.} We compare \sys{} with the following baselines. The OS baselines do not assume any logical index partitioning. The other baselines assume that the index is logically partitioned into  slices. Automatic NUMA balancing is
enabled for all systems.

\vspace{2pt}
\noindent1. \textbf{OS Default (OS:D)}. The OS handles data placement and core scheduling. In Linux, the default NUMA memory policy is \textit{local}, i.e., memory is allocated on the local NUMA node of the core that initiates the memory allocation request~\cite{numa-memory}. By default, a query can be executed on any of the cores. 

\vspace{2pt}
\noindent2. \textbf{OS Interleave (OS:I)}. The OS allocates memory in an interleaved fashion. The core scheduling follows the default OS policy. 

\vspace{2pt}
\noindent3. \textbf{Shared Everything-NUMA (SE:N)}~\cite{PorobicPBTA12}. Data is placed using a NUMA-aware policy, i.e., index slices with adjacent key ranges are allocated on the same NUMA node. The OS handles core scheduling.

\vspace{2pt}
\noindent4. \textbf{Shared Nothing-NUMA (SN:N)}~\cite{PorobicPBTA12}. The data placement follows the Shared Everything-NUMA strategy. Core scheduling maintains the data affinity of the index slice, i.e., a query can only be scheduled on one of the local cores where the data of the corresponding index slice resides.

\vspace{2pt}
\noindent5. \textbf{Shared Nothing-Thread (SN:T)}. Both  data placement and core scheduling are NUMA-aware following a Shared Nothing (SN) strategy~\cite{PorobicPBTA12}. A query can only be scheduled on the designated core reserved for the particular index slice that the query accesses. The schedules learned by \sys{} falls in this category.
    
5.1. \textbf{Grouped}. Index slices with adjacent key ranges are allocated in the same NUMA node and are scheduled in nearby cores. 
    
5.2. \textbf{Spread}. Index slices with adjacent key ranges are spread across all NUMA nodes.
    
5.3. \textbf{Mixed}. Following~\cite{PorobicPBTA12}, this Shared-Nothing strategy is a middle ground between the two strategies, Grouped and Spread.
    
5.4. \textbf{Random}. This Shared-Nothing strategy does not abide by any NUMA-aware heuristic. Rather, data placement and  core scheduling are done randomly generating a checkerboard pattern. 

\vspace{4pt}
\noindent\textbf{\sys{} Configurations}. 
\sys{}'s DT implementation is based on~\cite{LaiLTWH023,ChenLRLGLASM21}. We set the number of layers, attention heads, and embedding size to 6, 8, and 128, respectively. The context length of DT is set to 256, equaling the number of index slices. 
During inference on a target workload and a target server, the initial RTG of DT is set to 2$\times$ the best throughput observed in the fine-tuning dataset. During pre-training, \sys{'s} DT is trained on an NVIDIA A30 Tensor Core GPU for approximately 4 hours to reach 91.2\% accuracy on the offline dataset. In contrast, during fine-tuning, the base model is trained for up to 5 minutes on a fine-tuning dataset of up to 1900 samples, as it reaches an accuracy of over 90\%. Fine-tuning is relatively lightweight compared to pre-training, since most of the learning occurs during pre-training stage. The reported experiments in \S\ref{exp:machine} and \S\ref{exp:query} use a {\em single} round of fine-tuning. The inference takes up to 1.5 minutes on the same device.

\begin{figure*}[t]
\captionsetup[subfigure]{aboveskip=1pt}
\captionsetup[subfigure]{belowskip=2pt}
    \centering
    \begin{subfigure}[t]{0.95\columnwidth}
        \centering
        \includegraphics[width=\columnwidth]{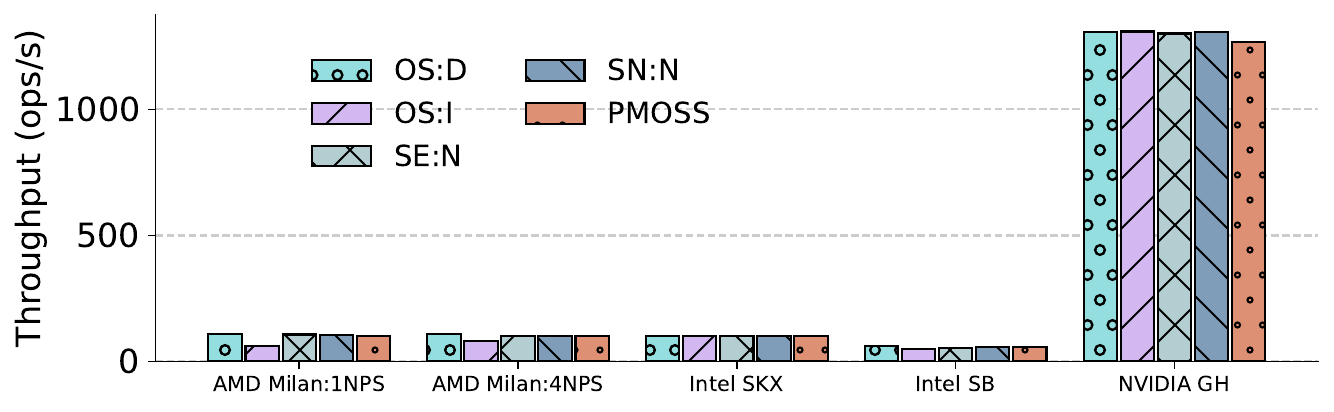}
        \caption{\sys{} vs OS baselines. 
        }
        \label{fig:ycsbe_linux}
    \end{subfigure}
    \begin{subfigure}[t]{0.95\columnwidth}
        \centering
       \includegraphics[width=\columnwidth]{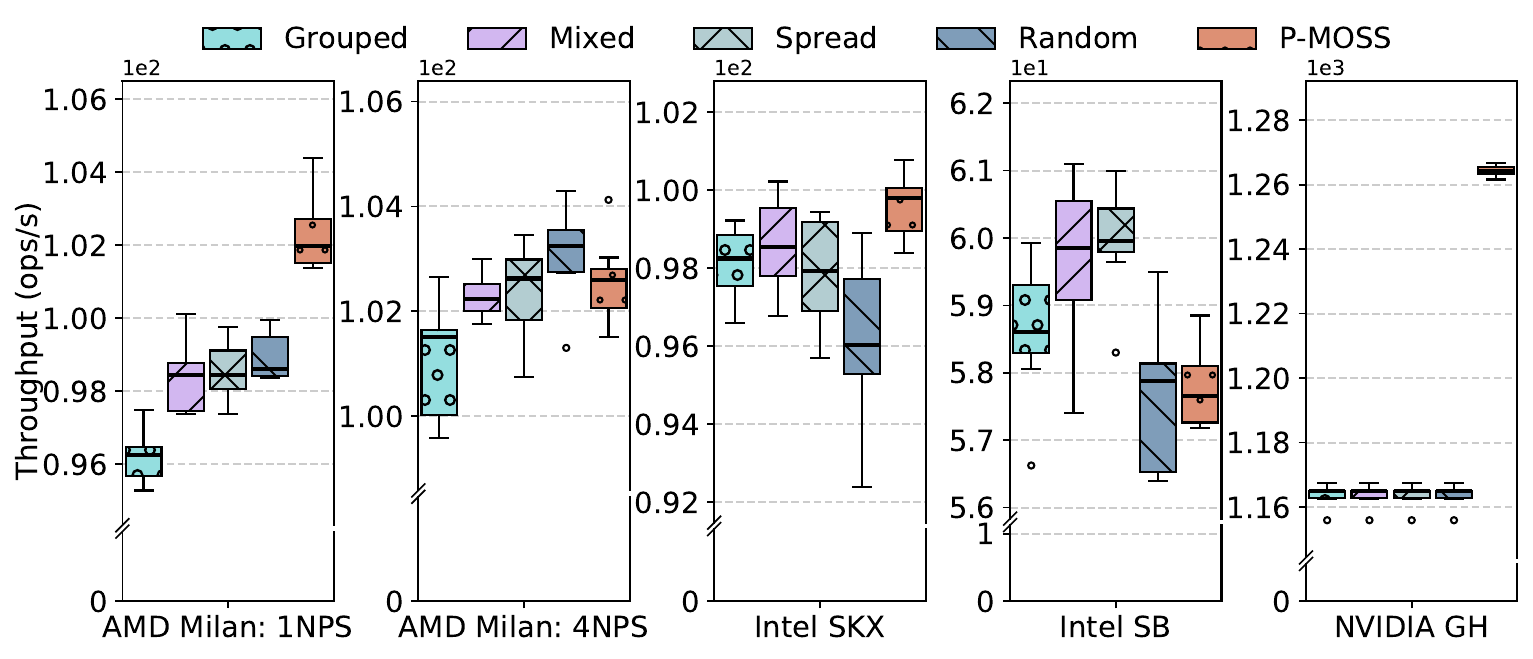}
        \caption{\sys{} vs SN:T baselines.
        }
        \label{fig:ycsbe_learned}
    \end{subfigure}
\captionsetup{
belowskip=-10pt, 
aboveskip=-0.1pt
}
\caption{\sys{} vs baselines on YCSB Scan workload.}
\label{fig:ycsbe}
\end{figure*}
\begin{figure*}[t]
\captionsetup[subfigure]{aboveskip=1pt}
\captionsetup[subfigure]{belowskip=2pt}
    \centering
    \begin{subfigure}[t]{0.95\columnwidth}
        \centering
        \includegraphics[width=\columnwidth]{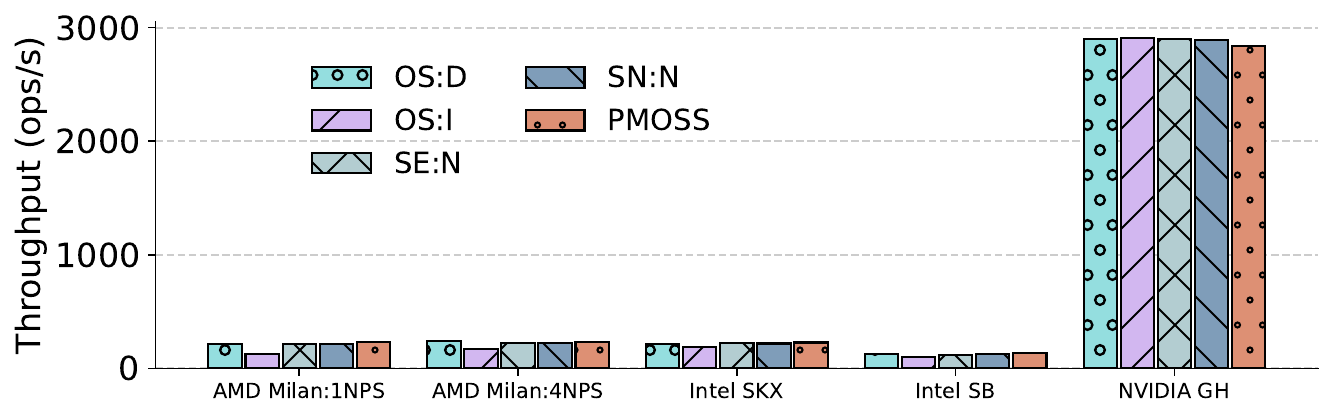}
        \caption{\sys{} vs OS baselines.}
        \label{fig:ycsbk2_linux}
    \end{subfigure}
    \begin{subfigure}[t]{0.95\columnwidth}
            \centering
        \centering 
         \includegraphics[width=\columnwidth]{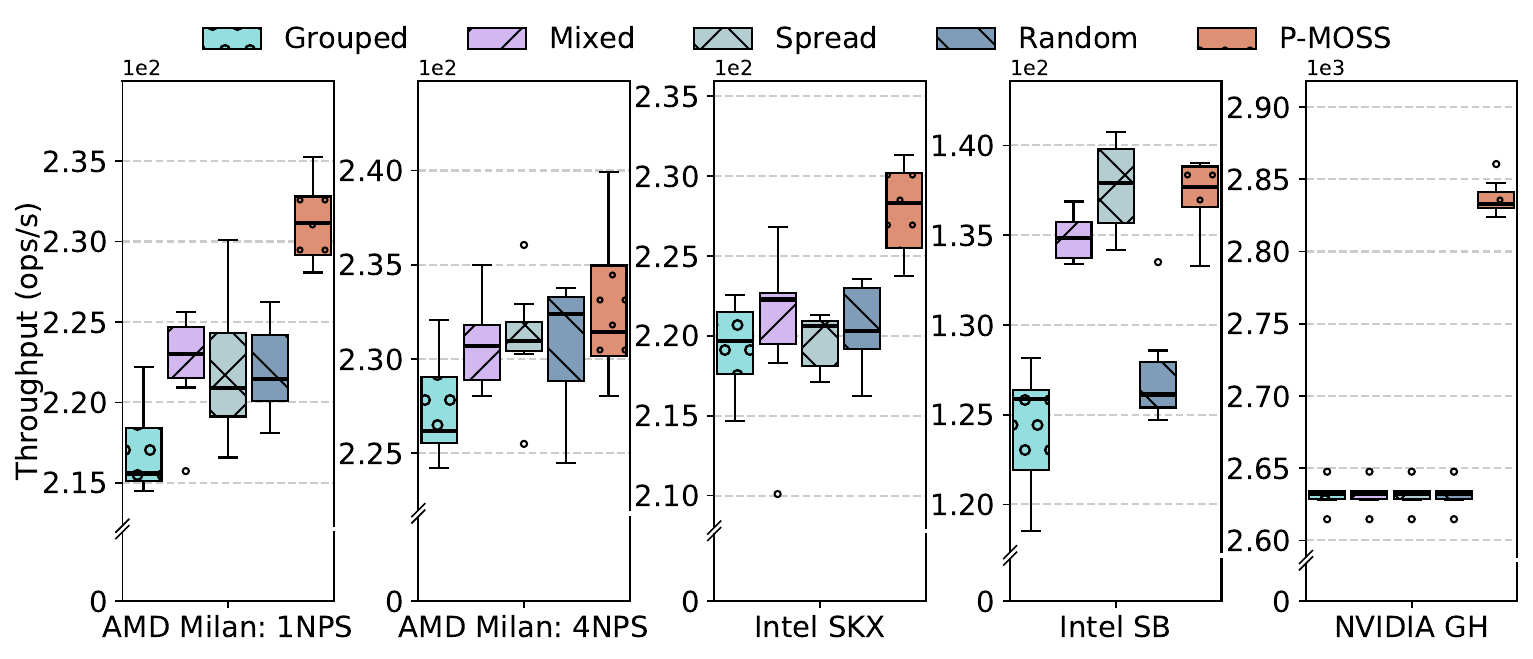}
        \caption{\sys{} vs SN:T baselines.}
        \label{fig:ycsbk2_learned}
    \end{subfigure}
\captionsetup{
belowskip=-10pt, 
aboveskip=-0.1pt
}
\caption{\sys{} vs baselines on YCSB Mixed workload.}
\label{fig:ycsbk2}
\end{figure*}
\subsection{Overall Performance}
\label{exp:machine}
We examine \sys{}'s performance  against the competing baselines for the B$^+$-Tree across all five testbed servers. The goal is to highlight the varying performance characteristics of the same scheduling strategy across different CPU vendors and NUMA architectures. We evaluate \sys{} on the YCSB benchmark~\cite{CooperSTRS10} with 1000M records. Each record has a 64-bit key and a 64-bit value. The query workload has 50\% reads and 50\% writes, and follows a Zipfian distribution with the Zipfian constant set to 0.99.

We start by showing \sys{}'s performance against the OS baselines (OS:D, OS:I) and NUMA-aware scheduling strategies (SE:N, SN:N) in Figure~\ref{fig:ycsba_linux}. Note that both SE:N and SN:N only place index data in a NUMA-aware manner. \sys{} outperforms all  approaches across all five servers. On average, \sys{'s} learned schedule has a speedup of $1.92\times, 1.90\times, 1.91\times, 3.66\times$ over the OS:D, OS:I, SE:N, SN:N strategies, respectively, across all five servers. Among the baselines, there is no single scheduling strategy that is dominant across servers, e.g., the interleaved strategy of OS (OS:I) performs  best for the AMD Milan (1NPS) and Intel Sandy Bridge server. However, the default OS strategy performs best for AMD Milan (4NPS) and the Intel Skylake X Server. For AMD servers, \sys{}'s performance gain  over the OS baselines can be attributed to a reduced number of data cache fills from a remote socket's cache and memory. Also, \sys{}  reduces the number of data cache fills from  external (off-chip) cache that is on the same socket. For the Intel servers, \sys{} reduces the number of remote memory accesses, significantly minimizing the latency of cache and memory stalls. Aside from the remote cache and memory accesses, \sys{}'s learned schedule also improves the DTLB cache behavior of the B$^+$-Tree and reduces the number of executed instructions.


The performance gap between \sys{} and  SE:N, SN:N  indicate how important the \textit{core scheduling policy} is, especially in the context of modern NUMA and Chiplet servers. The SN:N strategy yields $4.82\times, 3.12\times, 5.91\times, 3.08\times, 1.39\times$ performance degradation over \sys{} across the five testbed servers, respectively. This claim is further validated by the performance gap of 1.40$\times$ between \sys{} and the baselines for NVIDIA Grace Hopper that is a single-socket server with a single NUMA node. For the NVIDIA server, the only difference between the baselines and \sys{} lies in the core scheduling policy. All  approaches follow the same data partitioning technique, as it is a single socket server. By learning a better core scheduling policy, \sys{} improves the B$^+$-Tree's cache efficiency, and reduces the number of bus and memory accesses. This showcases the adaptive capability of \sys{}'s PMU-centric learned agent. 
Depending on the scenario, \sys{} is able to optimize different
hardware counters and learn better scheduling policies.

To further demonstrate \sys{'s} capability, we evaluate the performance of \sys{}'s learned schedule against different SN:T strategies, that take both core scheduling and data placement into account (cf. Figure~\ref{fig:ycsba_learned}). \sys{} 
dominates the baselines and outperforms the Grouped, Mixed, Spread and Random strategies by up to $3.09\%, 2.48\%, 3.27\%, 3.95\%$, respectively, across the five servers. 
Similar to the OS baselines, no single SN:T strategy exhibits the best performance across all the servers for the read-write workload. For example, the Random strategy yields better throughput in the AMD Milan: 1NPS processor. To the contrary, the Mixed strategy yields a better throughput in the AMD Milan: 4NPS and Intel Skylake X servers. 
The Grouped strategy yields a better throughput in the Intel Sandy Bridge server. Notice that the reported numbers for the random strategy is not a lower bound, as different random draws may yield outcomes worse than those reported in the paper.

\subsection{Query Workloads}
\label{exp:query}
This section examines \sys{} performance  for the B$^+$-Tree on three workloads: point lookup, scan and mixed workloads. The point lookup and the scan workloads consist of 100\% point lookups and range scans, respectively. Both workloads follow a zipfian distribution. The selectivity of each range scan is uniformly distributed within $[0.001\%,0.01\%]$ range. The mixed workload consists of 25\% reads, 50\% scans and 25\% inserts. The workload follows a Zipfian distribution. The selectivity of each range scan is uniformly distributed ranging up to 0.01\%. We evaluate \sys{} on point lookup, scan, and mixed workloads  in~\Cref{fig:ycsbc_linux,fig:ycsbc_learned},~\Cref{fig:ycsbe_linux,fig:ycsbe_learned}, and~\Cref{fig:ycsbk2_linux,fig:ycsbk2_learned}, respectively.

\vspace{4pt}
\noindent\textbf{Point Lookup Workload}: For point lookup, \sys{} outperforms the OS baselines and the SE:N, SN:N strategies by 2.57$\times$ on  average (cf. Figure~\ref{fig:ycsbc_linux}). Among the baselines, the SE:N strategy performs the best for AMD Milan: 1NPS and NVIDIA Server. In contrast, the OS:D strategy performs best for AMD Milan: 4NPS and Intel Skylake X server. The OS:I strategy performs best for the Intel Sandy Bridge server. This aligns with our previous observation of no one strategy being optimal for all the cases, i.e., machines or workloads. 

Among the SN:T strategies, \sys{'s} learned schedule   consistently delivers the best performance across all testbed servers (cf. Figure~\ref{fig:ycsbc_learned}). \sys{} has up to 14.39\% performance improvement over the SN:T strategies. \sys{} outperforms the Mixed strategy by 1.78\%, 1.58\%, 1.76\%, 4.99\%, 0.73\%, respectively, for the five servers. Aligned with our earlier observation for read-write workloads, the optimal baseline SN:T strategy differs across machines. The Mixed strategy is second to \sys{} in the AMD Milan: 4NPS server, while for the AMD Milan: 1NPS, Intel Skylake X and Intel Sandy Bridge servers, 
the Spread strategy is second to \sys{}.


\begin{figure*}[t]
\captionsetup[subfigure]{aboveskip=1pt}
\captionsetup[subfigure]{belowskip=2pt}
    \centering
    \begin{subfigure}[t]{0.95\columnwidth}
        \centering
        \includegraphics[width=0.95\columnwidth]{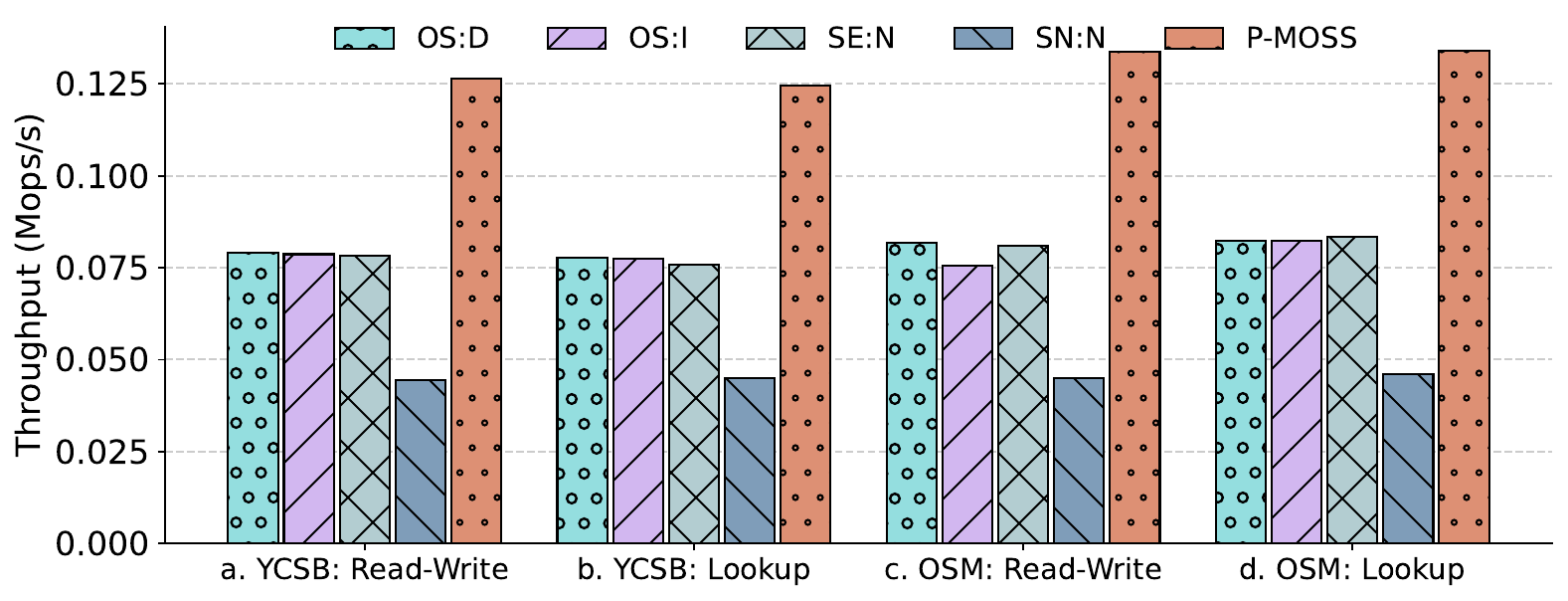}
        \caption{\sys{} vs OS 
        baselines.}
        \label{fig:ycsbu_linux}
    \end{subfigure}
    \begin{subfigure}[t]{0.95\columnwidth}
            \centering
        \centering 
         \includegraphics[width=\columnwidth]{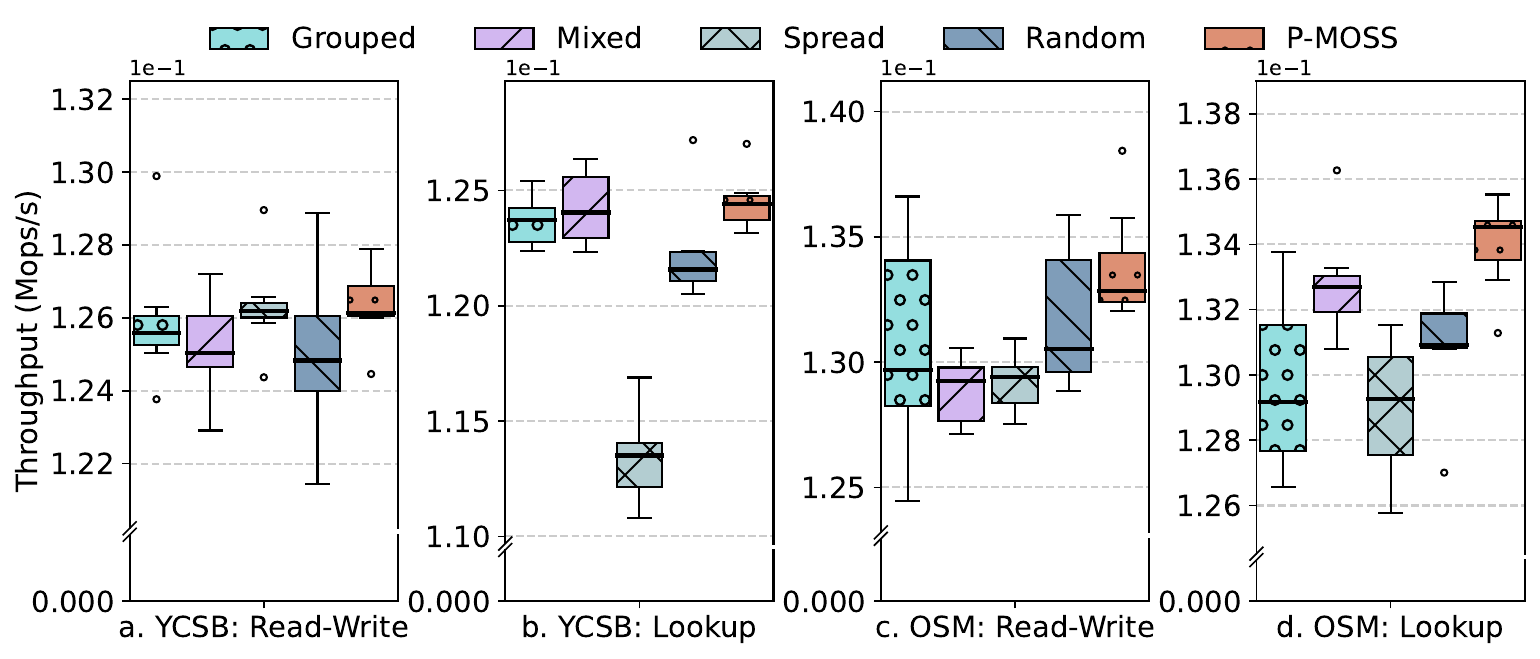}
    \caption{\sys{} vs SN:T baselines.}
    \label{fig:ycsbu_gen}
    \end{subfigure}
\captionsetup{
belowskip=-10pt, 
aboveskip=-0.1pt
}
\caption{\sys{} vs baselines on unseen 
    environment.}
\label{fig:ycsbu}
\end{figure*}
\vspace{4pt}
\noindent\textbf{Scan Workload}: 
For scan workload, \sys{} performs as good as the best performing baseline for the AMD and Intel servers. Compared to the read-write and point lookup workloads, the performance improvement is marginal. This is due to the low selective nature of the range scans, which can retrieve up to 10M records, often leading to cache thrashing. Thus, the impact of core scheduling policy is less prominent here. For the NVIDIA server, \sys{} experiences a minor performance regression. In this case, the  overhead of collecting hardware stats outweighs the benefit of the learned schedule. In contrast to the SN:T strategies, on 
average, \sys{} dominates the baselines by 4.94\% (cf. Figure~\ref{fig:ycsbe_learned}). \sys{} outperforms the Mixed strategy on the AMD Milan: 1NPS, AMD Milan: 4NPS, Intel Skylake and NVIDIA servers by 3.58\%, 0.37\%, 1.27\%, and 8.51\%, respectively. But, \sys{} shows a performance regression of 3.68\% on Intel Sandy Bridge server. 

\vspace{4pt}
\noindent\textbf{Mixed Workload}: On 
average, \sys{} outperforms the OS, SE:N, SN:N strategies by 1.15$\times$ (cf. Figure~\ref{fig:ycsbk2_linux}) for the mixed workload. As the mixed workload is scan dominated with 50\% range scans, the observations from the scan workload hold here as well. \sys{} outperforms the AMD and Intel servers by up to 1.33$\times$, and shows a minor performance regression for the NVIDIA server. Compared to the SN:T strategies, \sys{} demonstrates an average performance improvement of 3.70\% across all  servers. The Spread strategy performs best among the SN:T baselines for the AMD Milan: 4NP and Intel Sandy Bridge servers. This is in contrast to the read-write and point lookup workloads, where either Grouped or Mixed strategy performs best for these servers. For AMD Milan: 1NPS and Intel Skylake X servers, the Mixed strategy is second after \sys{}.


\subsection{Experimenting with Unseen Environments} 
\label{exp:unseen}
\begin{figure}[h]
    \captionsetup{aboveskip=1pt}
    \includegraphics[width=\columnwidth]{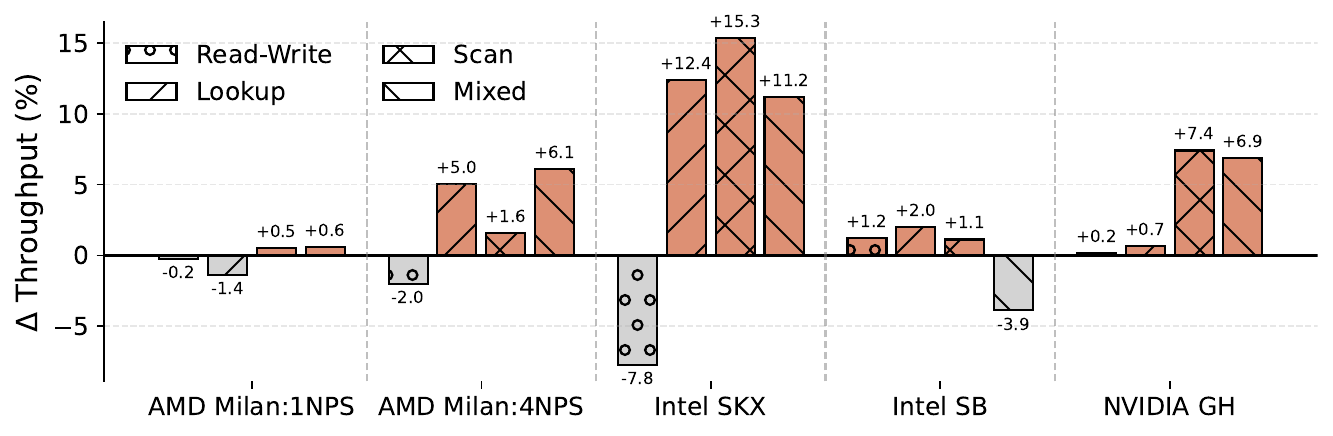}
    \caption{\sys{} vs the best SN:T baseline on unseen AMD, Intel and NVIDIA servers.}
    \label{fig:ycsbu_gen_more}
\end{figure}

We examine \sys{'s} performance in situations where the query workloads, data distribution or hardware platforms are not known beforehand. We test if \sys{}  generalizes beyond the data that it has been trained on and can keep up with the competing baselines. ~\Cref{fig:ycsbu_linux,fig:ycsbu_gen,fig:ycsbu_gen_more} gives \sys{'s} performance across multiple such cases. Notice that \sys{} does not log any scheduling policy in its offline dataset pertaining to the query workloads, datasets, or the hardware platforms studied below. 

\vspace{4pt}
\noindent\textbf{Hardware}. In this setting, we evaluate \sys{} on an IBM Power System S822LC server~\cite{ibm_power}. This is a dual socket 20 ($\times$8) core machine running Ubuntu 20.04 on a CloudLab~\cite{DuplyakinRMWDES19}  \verb|ibm8335| node with 255 GB of DDR4 RAM. Unlike the Intel, AMD, and NVIDIA servers that are based on the CISC architecture, the IBM server adopts a RISC architecture (PowerPC). Each physical core supports 8 simultaneous threads. Rather than integrating an on-die memory controller, it connects to a DRAM chip through dedicated `Centaur`` buffer chips. These buffer chips host a 16MB L4 cache. 

\vspace{4pt}
\noindent\textbf{Unseen Hardware Platforms.} 
Figures~\ref{fig:ycsbu_linux}a,~\ref{fig:ycsbu_linux}b and Figures~\ref{fig:ycsbu_gen}a,~\ref{fig:ycsbu_gen}b give \sys{'s} performance on the YCSB dataset for the 50\% read-write and 100\% point lookup workloads, respectively. Even though the offline dataset includes scheduling policies for the read-write and point lookup workloads, they originate from different servers. On average, \sys{} achieves 1.46$\times$ performance improvement over the competing baselines for both workloads. 

\vspace{4pt}
\noindent\textbf{Unseen Data Distribution.} 
Figures~\ref{fig:ycsbu_linux}c,~\ref{fig:ycsbu_linux}d and Figures~\ref{fig:ycsbu_gen}c,~\ref{fig:ycsbu_gen}d give \sys{'s} performance  on the OSM dataset~\cite{RamanHLA24,sosd-arxiv} for the 50\% read-write and 100\% point lookup workloads, respectively. The OSM dataset contains 800M records, and its data distribution differs from the YCSB benchmark. For the read-write and the point lookup workloads, on average, \sys{} outperforms the OS baselines 
by 2.01$\times$ and 1.94$\times$, respectively. 
Compared to the SN:T strategies, \sys{} achieves an average performance improvement of 2.41\%, 3.09\% for the respective workloads. 

Figure~\ref{fig:ycsbu_gen_more} gives additional experiments, where the  AMD, Intel and NVIDIA machines are the unseen hardware platforms. They are evaluated on  Read-Write, Lookup, Scan and Mixed YCSB workloads. On average, \sys{} achieves 2.8\% performance improvement over the best performing SN:T baselines across the respective hardware platforms and workloads. This showcases the generalization capability of \sys{}, and the importance of learning from diverse hardware platforms and workloads during the pre-training phase.


\subsection{Runtime System Components in \sys{}}
\begin{figure}[htbp]
    \captionsetup{belowskip=-12pt}
    \captionsetup{aboveskip=-0.5pt}
    \centering 
 \includegraphics[width=0.95\columnwidth]{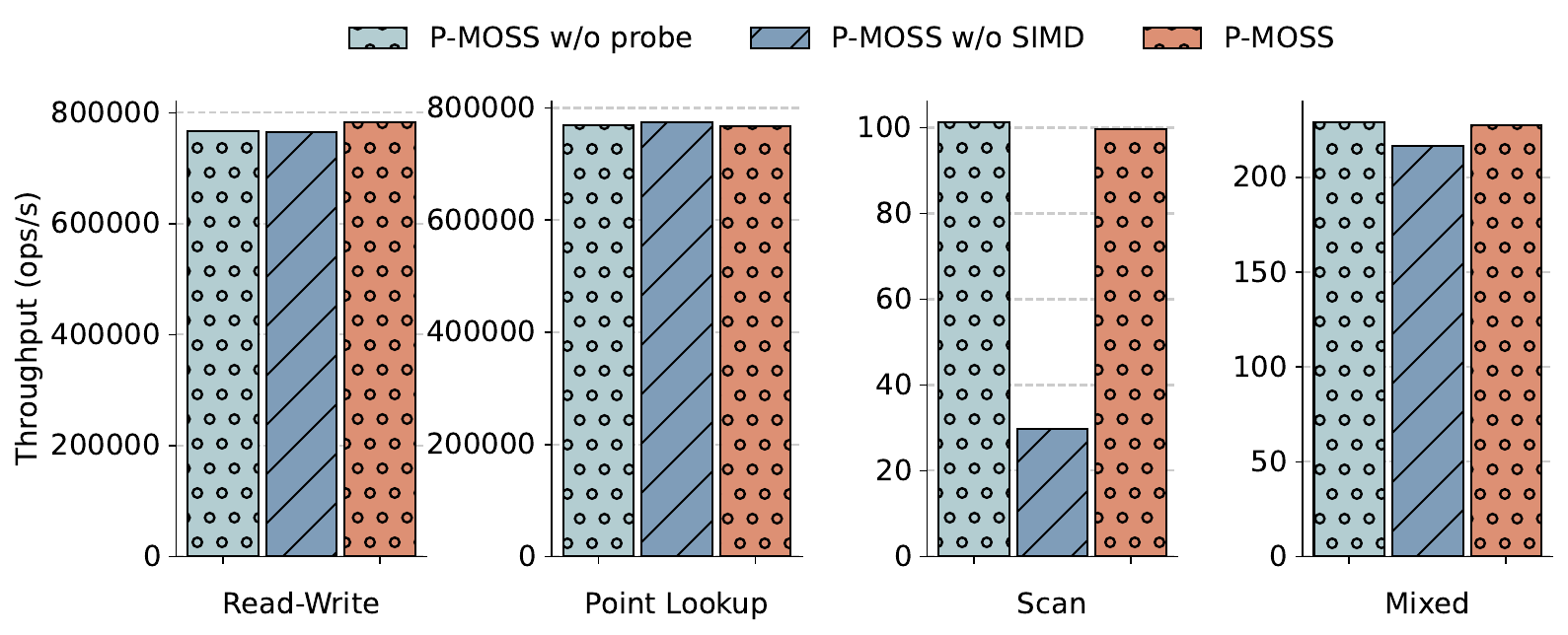}
    \caption{Analysis of \sys{}'s runtime system.}
    \label{fig:ycsb_sys}
\end{figure}

During all three stages: pre-training, post-training, and inference, \sys{} probes hardware PMU to profile the front-end, cache sub-system, memory sub-system and network interconnects. Figure~\ref{fig:ycsb_sys} shows the overhead associated with periodically probing the hardware PMUs in the Intel SKX server for the read-write, lookup, scan and mixed workloads. This includes profiling both the core and off-core statistics. On average, probing the PMU incurs a minimal 0.73\% performance degradation in \sys{}. During sweeping, \sys{} collects both core and off-core statistics from each worker core and off-core sweeper cores to generate the \verb|Hardware Snapshot| of the main-memory index under the current workload and scheduling policy. \sys{} uses extensive SIMD instructions to speed up this process. Figure~\ref{fig:ycsb_sys} details the advantage of using SIMD over scalar processing in this regard. On average, SIMD can improve \sys{}'s performance by 60.37\%. The performance improvement is more prominent for scan dominated workloads with low selectivity. Efficient SIMD processing of the PMU statistics (8$\times$ parallelism for AVX-512) enables \sys{} reduce the rate of cache pollution caused by the constant accumulation of the PMU statistics in the caches. For low selective range scans, this improves the cache hit rate as the effective cache capacity becomes larger.

\subsection{Learned Components in \sys{}}
\begin{figure}[htbp]
\captionsetup[subfigure]{aboveskip=-0pt}
\captionsetup[subfigure]{belowskip=1pt}
    \centering
    \begin{subfigure}[t]{0.52\columnwidth}
        \centering
        \includegraphics[width=\linewidth]{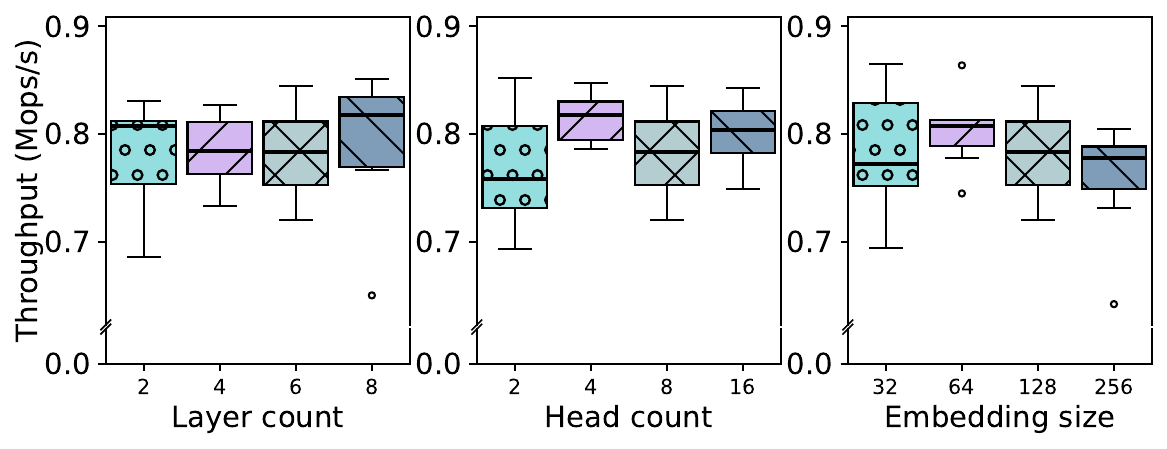}
        \caption{Ablation study of DT.}
        \label{fig:dt_ablation}
    \end{subfigure}
    \begin{subfigure}[t]{0.40\columnwidth}
        \centering
        \includegraphics[width=\linewidth]{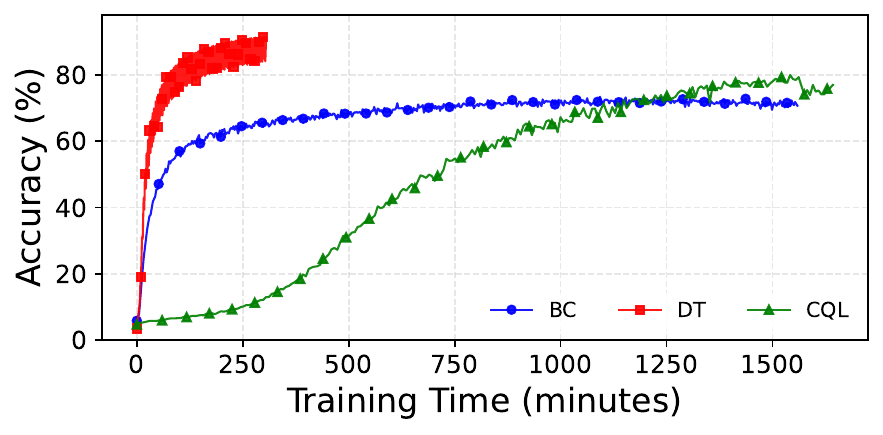}
        \caption{Training overhead.}
        \label{fig:altrl_train}
    \end{subfigure}
    \captionsetup{belowskip=-5pt, aboveskip=-0.5pt}
    \caption{Comparison of ablation results in \sys{}.}
    \label{fig:ablation_comparison}
\end{figure}

\noindent\textbf{Ablation Study of DT}. To study the impact of layer count, attention head count, and embedding size on \sys{'s} performance, we evaluate DT on the Intel SKX machine executing YCSB Read-Write workload. Refer to Figure ~\ref{fig:dt_ablation}. Increasing the layer and head count yields a better throughput, while it is the opposite for the embedding size. However, increasing any of these hyper-parameters raises the total number of model parameters, thereby increasing the training and inference overhead. Hence, \sys{}'s DT uses a default configuration of 6 layers, 8 attention heads, and an embedding size of 128.

\begin{figure}[htbp]
    \captionsetup{belowskip=-0pt}
    \captionsetup{aboveskip=1pt}
\includegraphics[width=0.95\columnwidth]{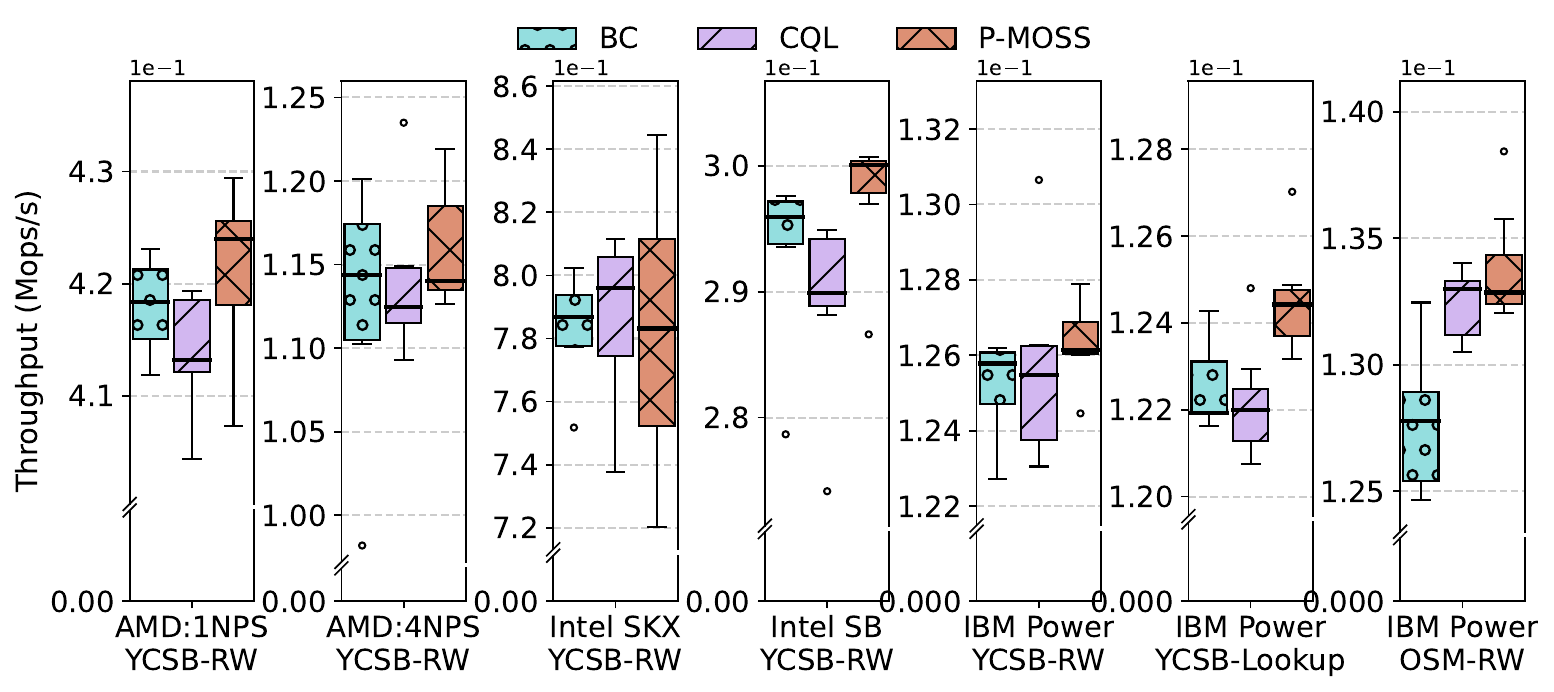}
    \caption{DT vs alternate Offline RL techniques in \sys{}. 
    }
    \label{fig:alt_rl}
\end{figure}

\noindent\textbf{Alternate Offline RL Techniques}. We compare the DT component of \sys{} with alternate Offline RL techniques, i.e., Behavior Cloning (BC) and Conservative Q-Learning (CQL). Figure~\ref{fig:altrl_train} shows the training overhead of DT compared to BC and CQL. Both BC and CQL achieve maximum accuracies of 71.1\% and 80\%, respectively, and incur high training overheads, requiring approximately 17 and 28 hours of training. In contrast, DT exceeds 90\% accuracy with only 4 hours of training due to the parallelization capability of the Transformer architecture itself. Note that all three algorithms use network architectures with a comparable number of parameters. Figure~\ref{fig:alt_rl} presents the performance of DT against BC and CQL across different hardware and workloads. DT outperforms BC and CQL by up to 1.82\% and 1.52\%, respectively. The performance gap becomes more prominent in unseen environment, i.e., the IBM machines, where DT outperforms the alternatives by up to 4.85\%.

\subsection{\sys{} Under Dynamic Environment}
\begin{figure}[htbp]
    \captionsetup{belowskip=-10pt}
    \captionsetup{aboveskip=1pt}
\includegraphics[width=0.95\columnwidth]{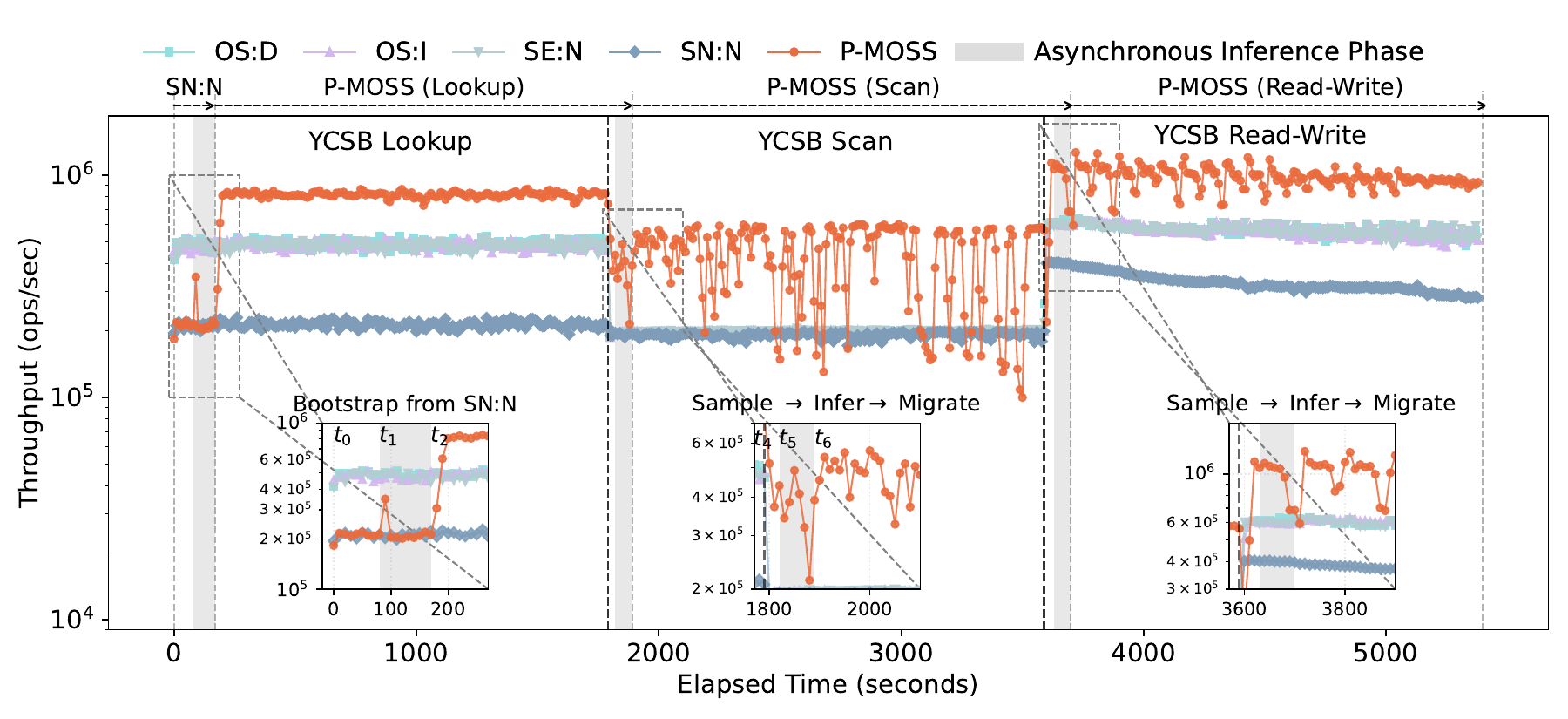}
    \caption{\sys{} under dynamic environment.}
    \label{fig:pmoss_dynamic}
\end{figure}

Figure~\ref{fig:pmoss_dynamic} shows the life cycle of \sys{} as it adapts to transitions in the YCSB workload: Lookup $\rightarrow$ Scan $\rightarrow$ Read-Write on the Intel Skylake X server. At Time $t_0$, P-MOSS bootstraps the B$^+$-Tree (initialized with 500M records) with the Linux SN:N strategy, and generates the Hardware Snapshots of the B$^+$-Tree under the SN:N strategy. At Time $t_1$, \sys{} delegates the inference task to a GPU along with the collected Hardware Snapshots to infer a new scheduling policy for the current Lookup workload. Notice that the inference is performed asynchronously, i.e., \sys{} continues to execute queries with the current SN:N strategy until the new schedule is generated. At Time $t_2$, once the inference completes, \sys{} moves pages using the Linux \texttt{move\_pages} system call to establish the new scheduling policy. When the workload later shifts to Scan and Read-Write, \sys{} repeats this process of collecting hardware snapshots, performing inference, and migrating pages accordingly. \sys{} generates a scheduling policy asynchronously {\em once} for the entire query workload, i.e., the scheduling policy generated at Time $t_3$ continues until the next workload transition at Time $t_6$. Notice that  migration occurs concurrently with query execution as the migration of the index slices are offloaded to the worker cores. 
Across the entire workload span, \sys{} outperforms the OS:D, OS:I, SE:N, SN:N strategies by 1.72$\times$, 1.73$\times$, 1.73$\times$, and 2.94$\times$, respectively. For individual workloads, \sys{} achieves up to 1.53$\times$, 2.18$\times$, and 1.71$\times$ better performance than the best-performing baseline for the Lookup, Scan, and Read-Write workloads, respectively.

\section{Lessons Learned}
\noindent 1. Spatial scheduling is essential for efficiently utilizing modern hardware platforms to maximize main-memory index performance. A consistent trend observed throughout all the experiments is that \sys{} outperforms the OS baselines, i.e., OS:D and OS:I, highlighting the importance of data partitioning. It also dominates the SE:N, SN:N and SN:T strategies highlighting the importance of core scheduling besides data partitioning to improve main-memory index performance in NUMA and Chiplet servers. 

\vspace{2pt}
\noindent 2. The design space for scheduling a main-memory index on modern hardware platforms is 
combinatorial in nature. No single heuristic dominates across this vast solution space, as evident from our experiments. Different hardware and different query workloads react differently to different scheduling policies. In such uncertain scenarios, Machine Learning techniques, particularly those inspired by LLMs (e.g., Next Token Prediction) are well suited to navigate the large solution space and learn better scheduling policies specific to the target hardware and workload patterns.

\vspace{2pt}
\noindent 3. Formulating optimization tasks, e.g., scheduling, as a Next Token Prediction problem enables the adoption of Offline Reinforcement Learning. The benefits are two fold. First, it allows \sys{} to leverage large datasets and powerful models, e.g., GPT, in a data-driven manner that  has been at the core of the LLMs' success. Second, this completely isolates the learning process from the database critical path, and allows systems to focus solely on learning scheduling decisions that can adapt across diverse hardware configurations and query workload patterns.

\vspace{2pt}
\noindent 4. The performance statistics from hardware PMUs are a strong candidate for feature representation in Machine Learning techniques for scheduling and Machine Learning for Databases (ML4DB, for short) optimization tasks. Due to the non-determinism and the high dimensionality of these statistics, large datasets and large models, e.g., Transformers, are essential to effectively leverage these hardware statistics, and vice versa. Hardware statistics can provide ample data to train large ML models for database optimization tasks, as observed in \sys{}.

\section{Conclusion}
We present \sys{}, a Performance Monitoring Unit (PMU)-driven learned scheduling framework that improves query processing in main-memory indexes on modern multi-core NUMA and Chiplet servers. Drawing inspiration from LLMs, \sys{} frames scheduling as a Next Token Prediction task, allowing it to adopt Offline RL paradigm. Combined with the hardware performance statistics from PMU, this formulation allows the learning process of \sys{} scale across large datasets and model sizes, resulting in better schedules that adapt effectively across diverse hardware architectures and workload patterns. Experiments on the B$^+$-Tree in diverse settings illustrate that \sys{} improves index performance up to $6 \times$. Overall, \sys{'s} performance results suggest that foundational LLM concepts, e.g., Next Token Prediction 
can effectively guide DBMS optimization tasks. 
While this paper is presented in the context of spatial scheduling over main-memory indexes, the techniques in this paper can be generalized beyond learning spatial scheduling policies to designing DBMS optimizations.

\bibliographystyle{ACM-Reference-Format}
\bibliography{sample}

\end{document}